\documentclass[aps,twocolumn,superscriptaddress,showpacs,article,prb,nofootinbib,longbibliography]{revtex4-2}

\usepackage[T1]{fontenc}
\usepackage{color}
\usepackage{verbatim}
\usepackage{amsmath}
\usepackage{bm}
\usepackage{amssymb}
\usepackage{graphicx}
\usepackage{esint}
\usepackage{esvect}
\usepackage{braket}
\usepackage{soul}
\usepackage{amsmath,bm}
\usepackage{graphicx}
\usepackage{subfigure}
\usepackage{url}
\usepackage{lipsum}
\usepackage[dvipsnames]{xcolor}
\usepackage{hyperref}
\usepackage[resetlabels]{multibib}
\usepackage[normalem]{ulem}

\makeatletter
\@ifundefined{textcolor}{}
{%
 \definecolor{BLACK}{gray}{0}
 \definecolor{WHITE}{gray}{1}
 \definecolor{RED}{rgb}{1,0,0}
 \definecolor{GREEN}{rgb}{0,1,0}
 \definecolor{BLUE}{rgb}{0,0,1}
 \definecolor{CYAN}{cmyk}{1,0,0,0}
 \definecolor{MAGENTA}{cmyk}{0,1,0,0}
 \definecolor{YELLOW}{cmyk}{0,0,1,0}
}

\usepackage{mathrsfs}

\draft
\makeatother

\begin{document}

\title{Magnetization amplification in the interlayer pairing superconductor 4Hb-TaS$_2$}

\author{Chunxiao Liu}
\affiliation{Department of Physics, University of California,  Berkeley, CA 94720, USA}
\author{Shubhayu Chatterjee}
\affiliation{Department of Physics, University of California, Berkeley, CA 94720, USA}
\affiliation{Department of Physics, Carnegie Mellon University, Pittsburgh, PA 15213, USA}
\author{Thomas Scaffidi}
\affiliation{Department of Physics, University of California, Irvine, Irvine, CA 92697, USA}
\affiliation{Department of Physics, University of Toronto, 60 St. George Street, Toronto, Ontario, M5S 1A7, Canada}
\author{Erez Berg}
\affiliation{Department of Condensed Matter Physics, Weizmann Institute of Science, Rehovot 76100, Israel}
\author{Ehud Altman}
\affiliation{Department of Physics, University of California,  Berkeley, CA 94720, USA}
\affiliation{Materials Sciences Division, Lawrence Berkeley National Laboratory, Berkeley, CA 94720, USA}

\begin{abstract}

A recent experiment on the bulk compound 4Hb-TaS$_2$ reveals an unusual time-reversal symmetry-breaking superconducting state that possesses a magnetic memory not manifest in the normal state. Here we provide a mechanism for this observation by studying the magnetic and electronic properties of 4Hb-TaS$_2$. We discuss the criterion for a small magnetization in the normal state in terms of spin and orbital magnetizations. Based on an analysis of lattice symmetry and Fermi surface structure, we propose that 4Hb-TaS$_2$ realizes superconductivity in the interlayer, equal-spin channel with a gap function whose phase winds along the Fermi surface by an integer multiple of $6\pi$. The enhancement of the magnetization in the superconducting state compared to the normal state can be explained if the state with a gap winding of $6\pi$ is realized, accounting for the observed magnetic memory. 
We discuss how this superconducting state can be probed experimentally by spin-polarized scanning tunneling microscopy.

\end{abstract}

\maketitle

\section{Introduction}
 
\begin{figure}
\centering
\includegraphics[width=0.48\textwidth]{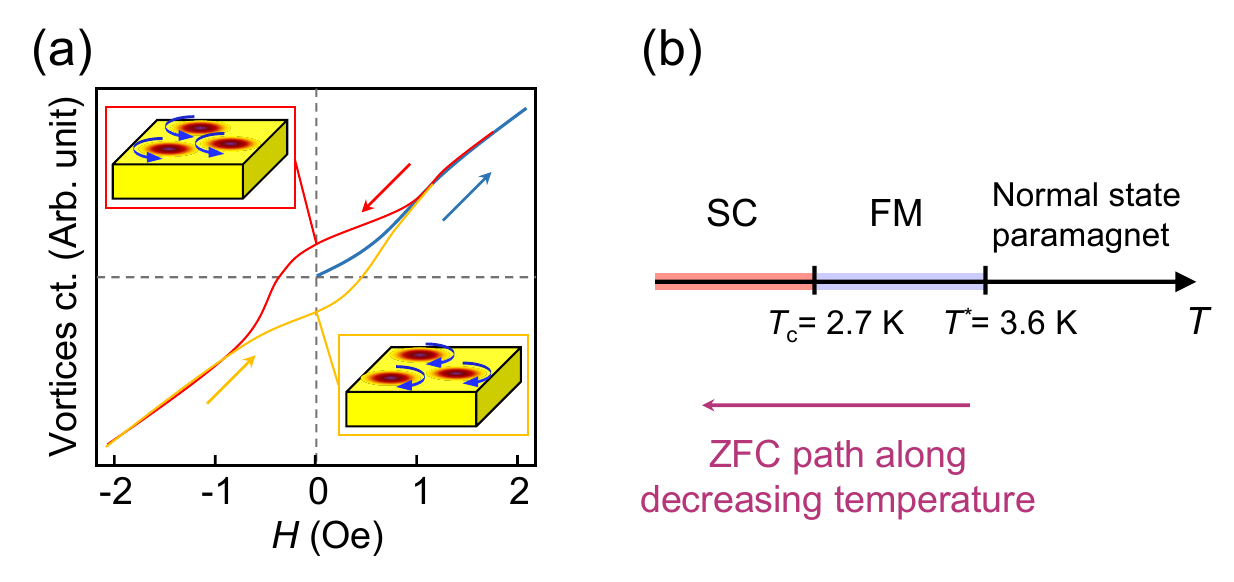}
\caption{(a) A reproduction of the hysteresis curve from the magnetometry experiment \cite{persky2022magnetic}. The vertical axis counts the number of spontaneous vortices, and the horizontal axis is the training field applied above $T_c$; the insets illustrate the spontaneous vortices. (b) A heuristic phase diagram along the temperature axis inferred from the training-ZFC process. The magenta arrow defines the ZFC path.}\label{Phase_diagram}
\end{figure}

A recent experiment raises an intriguing puzzle about the interplay between magnetism and superconductivity in the multilayer transition metal dichalcogenide compound 4Hb-TaS$_2$ \cite{persky2022magnetic}. As expected, the sample exhibits vortices when cooled in a magnetic field below the superconducting $T_c\!=\!2.7$ K and no vortices appear below $T_c$ for zero-field-cooling (ZFC). 
However, the behavior of the system during a mixed training-ZFC protocol poses a puzzle.
Specifically, vortices appear spontaneously if the system is cooled in zero field after being trained in a magnetic field applied  \textit{above} the superconducting $T_c$ and below $T^*\!=\!3.6$ K  although there is no direct sign of a residual magnetization above $T_c$. 
This surprise is best reflected in the hysteresis curve for the vortex density versus the training field applied above $T_c$, as shown in Fig.~\ref{Phase_diagram}(a).

The origin of the spontaneous vortices that appear below $T_c$ is not understood. However the fact that the vortex density and chirality respond to a training field applied above $T_c$ sets important constraints on the possibilities. In particular it implies that a state with spontaneously broken time-reversal symmetry (TRS) must already have been established above $T_c$, but the magnetization in this state is too small to be detected by the SQUID magnetometer. These remarkable observations raise a natural question: how can a small magnetization in the parent metallic phase be highly amplified in the descendant superconductor?

One possibility for the time-reversal symmetry-breaking (TRSB) state proposed previously is a chiral spin liquid (CSL) or chiral metallic state on the 1T layers \cite{persky2022magnetic,lin2022kondo}. The monolayer compound 1T-TaS$_2$ is known to be a Mott insulator \cite{wilson1975charge,
PhysRevLett.73.2103,
perfetti2006time}. If the 1T layers in 4Hb-TaS$_2$ are in a CSL phase, the spin chirality can carry the memory of the training field without generating a detectable magnetization, hence orienting a chiral superconductor below $T_c$ \cite{persky2022magnetic,lin2022kondo}. Light doping of the Mott insulators due to charge transfer to the 1H layers in 4Hb-TaS$_2$ \cite{wang2018surface,PhysRevB.102.075138,nayak2021evidence} may turn the CSL into a chiral metal, with similar effect. However ab initio calculation and spectroscopic experiments indicate an almost completely depleted 1T band \cite{nayak2021evidence,nayak2023first}. These results call for an alternative explanation of the memory effect not relying on lightly doped 1T layers.

In this paper we investigate a mechanism for the magnetic memory observed in the superconducting state that is consistent with the structure and symmetry of 4Hb-TaS$_2$. 
Our mechanism rests on the assumption
that the metallic state above $T_c$ hosts at least a weak ferromagnetic (FM) moment, which may be too small to be detected (see Fig. ~\ref{Phase_diagram}(b)). We then determine the superconducting instabilities consistent with such a normal state and examine the requirements on the SC states for enhanced magnetization. 
Our key insight is that, the TRSB order, while suppressing the intralayer conventional BCS pairing, may favor an interlayer equal-spin pairing state protected by inversion symmetry. An imbalance between the spin-up and down pairings generically happens in this pairing state, such that the minority spin component can remain in an unpaired normal state. The angular momentum carried by the majority Cooper pairs gives rise to enhanced magnetization in the SC phase. The proposed mechanism leads to a number of predictions including a spin dependent ``partial gap'' structure, a linear-in-temperature specific heat, and the possibility of a second transition temperature below $T_c$, all of which are consistent with existing experiments \cite{doi:10.1126/sciadv.aax9480,nayak2021evidence} or testable in the future.

{The paper is structured as follows. In Sec.~\ref{sec:model} we introduce lattice structure and symmetry of 4Hb-TaS$_2$ and present a tight-binding model for the bands near the Fermi surface. In Sec.~\ref{sec:mag_normal_state} we calculate the magnitude of magnetization in presence of weak ferromagnetism in the proposed TRSB normal state, and discuss constraints on the strength of the magnetic order from the experimental observations. We then analyze the pairing channels allowed by the Fermi surface geometry and lattice symmetries in Sec.~\ref{sec:two_pairing_channels}, from which we propose that the material favors an interlayer pairing. This interlayer pairing is then examined in great detail: we classify in Sec.~\ref{Sec:Interlayer pairing} the symmetries of the interlayer pairing, and point out in Sec.~\ref{sec:consequence_topology_unpair} two major physical consequences of the interlayer pairing -- nontrivial gap winding and imbalance in the spin up and spin down pairing sectors. In Sec.~\ref{sec:mag}, we calculate the magnetization for several interlayer superconducting states, one of which exhibits major enhanced magnetization that match the experimental value. We conclude the paper in Sec.~\ref{sec:con_dis} with a summary of results and discussions for future experimental verification of our predictions.}


\section{Model Setup}\label{sec:model}

\subsection{Lattice symmetry}

The bulk 4Hb-TaS$_2$ structure is shown in Fig.~\ref{fig:lattice}(a).
The lattice is centrosymmetric,  with inversion centers residing on Ta atoms in $\mathrm{T}$ layers. The inversion $i$ interchanges the $\mathrm{H}$ and $\mathrm{H'}$ layers, which, as we will show later, is crucial to the formation of interlayer superconductivity.

The lattice has the following symmetries: $m_z$ is reflection with horizontal mirror plane in the 1$\mathrm{H}$ layer; $m_x$ is reflection with vertical mirror plane that contains the bond $R_1$ (see Fig.~\ref{fig:lattice}(b)); $s_2$ is a two-fold screw along the vertical direction, and $c_{3z}$ is a threefold rotation along the vertical direction. 

ARPES data \cite{doi:10.1126/sciadv.aax9480} suggests that the bands near the Fermi energy come from the H and $\mathrm{H'}$ layers and consist of three orbitals $|d_{z^2}\rangle$, $|d_{x^2-y^2}\rangle$, $|d_{xy}\rangle$ and two spins $\sigma\!=\uparrow,\downarrow$ \cite{liu2013three}. In this orbital subspace the dominant spin--orbit coupling (SOC) is the spin $S_z$-preserving Ising SOC $\hat{L}_z\hat{S}_z$.

\subsection{Tight-binding model}

\begin{figure}
\centering
\includegraphics[width=0.48\textwidth]{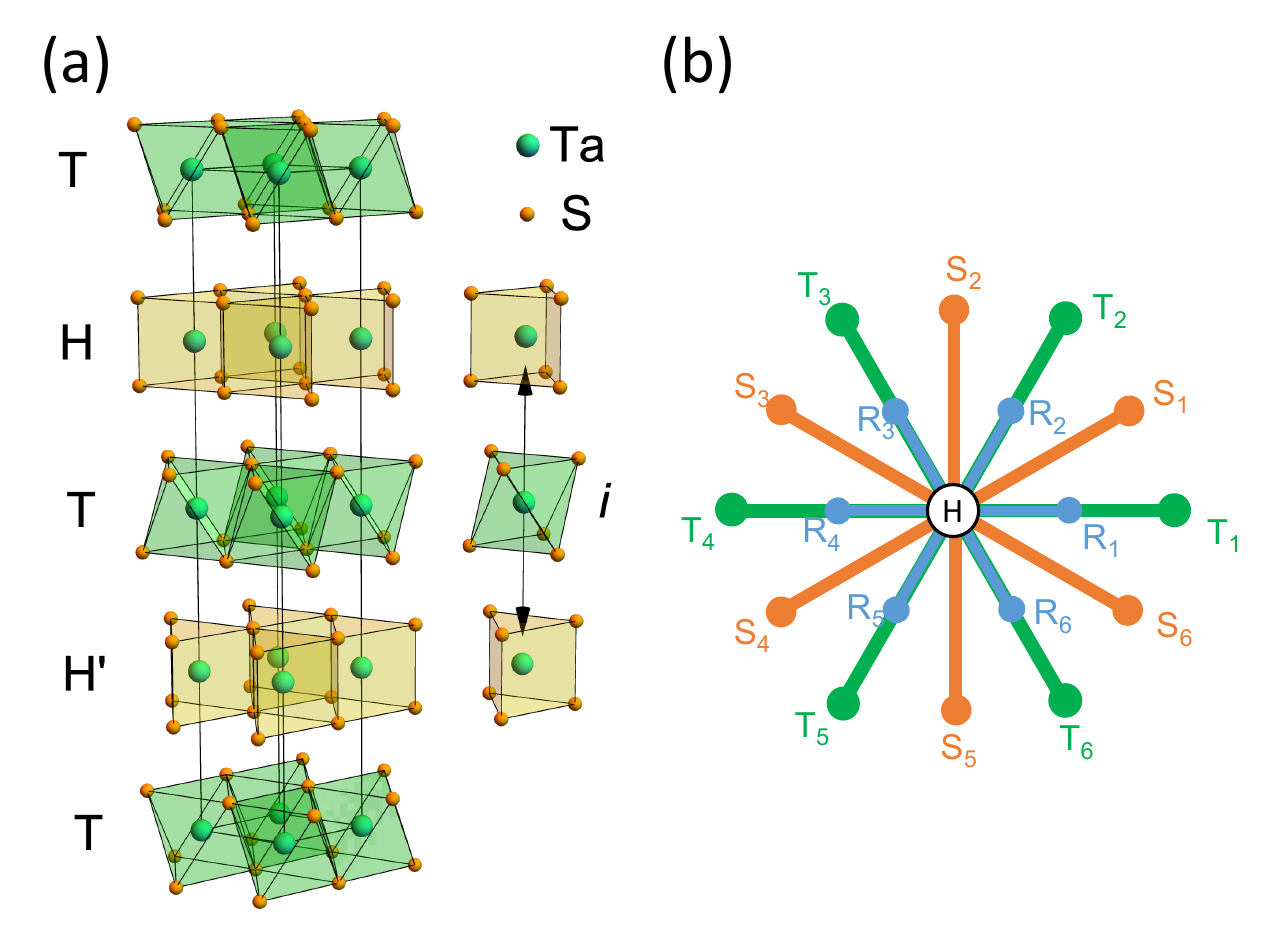}
\caption{Lattice information of 4Hb-TaS$_2$. (a) The 3D 4Hb-TaS$_2$ lattice. The tantulum atoms form a simple stacking of triangular lattices with a four-layer-periodic unit cell  $\mathrm{T}$--$\mathrm{H}$--$\mathrm{T}$--$\mathrm{H'}$, where $\mathrm{T}$ and $\mathrm{H}$ denote two different layer structures. The $\mathrm{H}$ and the $\mathrm{H'}$ layers are related by inversion symmetries $i$ with inversion ceters on Ta atoms of the $\mathrm{T}$ layers. (b) Definition of the nearest, 2nd-nearest, and 3rd-nearest neighbor bonds within the $\mathrm{H}$ layer.}\label{fig:lattice}
\end{figure}

For all our microscopic calculations in later sections we employ a six-band tight-binding model derived from DFT calculations \cite{MARGALIT2021168561}  that matches ARPES data \cite{doi:10.1126/sciadv.aax9480}. Define 
\begin{equation}\label{Ddbasis2}
d^\dag_{\bm{r},\mathrm{H}}
=(d^\dag_{\bm{r},\mathrm{H},\uparrow},d^\dag_{\bm{r},\mathrm{H},\downarrow}),
\end{equation}
where $\bm{r}$ labels lattice sites on one layer, $\sigma=\uparrow,\downarrow$ labels spins, and
\begin{equation}\label{ddbasis}
d^\dag_{\bm{r},\mathrm{H},\sigma} = \left(d^\dag_{z^2,\sigma,\bm{r},\mathrm{H}},d^\dag_{xy,\sigma,\bm{r},\mathrm{H}},d^\dag_{x^2-y^2,\sigma,\bm{r},\mathrm{H}}\right)
\end{equation}
is the creation operators for the orbitals $|d_{z^2}\rangle$, $|d_{x^2-y^2}\rangle$, and $|d_{xy}\rangle$ and spins $\sigma$ at site $\bm{r}$.
The Hamiltonian for the $\mathrm{H}$ layer is, in the Fourier-transformed momentum space,
\begin{equation}\label{tbnsH}
H_{\text{H}} = \sum_{\bm{k}} d^\dag_{\bm{k},\mathrm{H}} \mathcal{H}_{\mathrm{H}}(\bm{k}) d_{\bm{k},\mathrm{H}},
\end{equation}
with
\begin{equation}
\mathcal{H}_{\mathrm{H}}(\bm{k})
=E_0 + \sigma_0\otimes \left(\sum_{i=1}^6 R_i e^{i \bm{R}_i\cdot \bm{k}} + S_i e^{i \bm{S}_i\cdot \bm{k}}
+T_i e^{i \bm{T}_i\cdot \bm{k}}\right),
\end{equation}
where $\sigma_0$ is the identity matrix in the spin space, $R_i$, $S_i$, and $T_i$ are $3\times 3$ matrices consisting of nearest neighbor, next-nearest neighbor and third neighbor hoppings (see Fig.~\ref{fig:lattice}(b)). The onsite term is 
\begin{equation}
E_0 = \sigma_0 \otimes \text{diag}(
\epsilon_0-\mu_0, \epsilon_1-\mu_0 ,\epsilon_2-\mu_0)+  \frac{\lambda_{\text{SO}}}{2}\sigma^z \otimes L^z,
\end{equation}
with $\sigma^z$ the $z$-component Pauli matrix and $L^z  = \left(\begin{smallmatrix}0&0&0\\0&0&2i \\ 0&-2i&0\end{smallmatrix}\right)$. 
The values of the hopping matrices and onsite energies are give in Appendix \ref{app:tb}.

As a consequence of the Ising SOC which conserves the spin $\sigma$, the eigenstates are decoupled into spin-up and spin-down sectors
\begin{equation}
H_{\mathrm{H}} = \sum_\sigma \sum_{n=1,2,3} E_{n,\sigma,\bm{k}}c^\dag_{n,\sigma,\bm{k}}c_{n,\sigma,\bm{k}},
\end{equation}
where the creation operators $c^\dag_{n,\sigma,\bm{k}}$ are related to the orbital operators by $c^\dag_{n,\sigma,\bm{k}} = \sum_{\ell=z^2,xy,x^2-y^2} u^{(\ell)}_{n,\sigma,\bm{k}} d^\dag_{\ell,\sigma,\bm{k},\mathrm{H}}$, where $u_{n,\sigma,\bm{k},\mathrm{H}}$ are vectors that diagonalize the Hamiltonian matrix $\mathcal{H}_{\mathrm{H}}(\bm{k}) u_{n,\sigma,\bm{k},\mathrm{H}} = E_{n,\sigma,\bm{k}}u_{n,\sigma,\bm{k},\mathrm{H}}$. The dispersion $E_{n,\sigma,\bm{k}}$ along high symmetry paths in the Brillouin zone (BZ) is shown in Fig.~\ref{fig:band}(a). 

The Hamiltonian for the $\mathrm{H'}$ can be obtained from that for the $\mathrm{H}$ layer via inversion $i$:
\begin{equation}\label{relationbetweenlayer}
\mathcal{H}_{\mathrm{H}'}(\bm{k}) =\mathcal{H}_{\mathrm{H}}(-\bm{k}).
\end{equation}

\section{Magnetization in the normal state}\label{sec:mag_normal_state}

\subsection{Breaking of time reversal symmetry in the normal state}

The ability to train the vortex state using a field below $T^*\!=\!3.6$ K implies a TRSB order with a spontaneous magnetization in the normal state. Here we leave open its microscopic origin, but assume that the corresponding TRSB order parameter, $\phi$, couples to the electrons as a Zeeman field. In addition to spin polarization, this coupling leads to orbital magnetization in the form of bond currents through the Ising SOC.
We expect that the TRSB order has multiple frozen FM domains with random orientations, which get realigned by the training field below $T^*$. Below we estimate the magnetization of these aligned domains
and discuss under which conditions it may be very small in the normal state and amplified below $T_c$.

\subsection{Estimation of the weak ferromagnetism}

Since we assumed that $\phi$ couples to the electrons as a Zeeman field, we introduce an effective field $B_{\text{eff}}$ to measure the splitting between the spin-up and -down bands: $E_\uparrow - E_\downarrow \simeq \phi\simeq\mu_B B_{\text{eff}}$. Below we calculate the out-of-plane magnetization, $M_{\text{tot}}$ induced by $\phi$ (or equivalently, $B_{\text{eff}}$). The total magnetization is
\begin{equation}
M_{\text{tot}} =  M_{\text{spin}} + M_{\text{orb}},
\end{equation}
the first term is the spin magnetization
\begin{equation}
M_{\text{spin}}= \frac{g_s}{2} \frac{\mu_B}{V_{\text{2D unit cell}}}  \sum_n \int \frac{d^2k}{(2\pi)^2} c^\dag_{n,\bm{k}} \sigma^z c_{n,\bm{k}},
\end{equation}
with $g_s\approx 2$ the spin $g$-factor, $V_{\text{2D unit cell}}$ the area of the unit cell in a single $\mathrm{H}$ layer, and we defined $c^\dag_{n,\bm{k}} = (c^\dag_{n,\uparrow,\bm{k}},c^\dag_{n,\downarrow,\bm{k}})$. The second term is orbital magnetization, given by \cite{PhysRevB.74.024408}
\begin{equation}\label{normalorbmag}
\begin{aligned}
&M_{\text{orb}} = \frac{|e|}{\hbar} \times \\
&\sum_n \int \frac{d^2{k}}{(2\pi)^2}
\mathrm{Im}\langle \partial_{k_x} u_{n\bm{k}}| H_{\bm{k}} + E_{n \bm{k}}-2 E_{\text{F}}|\partial_{k_y} u_{n\bm{k}}\rangle f_{n\bm{k}}.
\end{aligned}
\end{equation}
when time-reversal symmetry breaking is weak, the total magnetization $M_{\text{tot}}$ is linearly proportional to the effective symmetry breaking field, $B_{\text{eff}}$. Express $B_{\text{eff}} = \mathsf{B}_{\text{eff}}$ T, where $\mathsf{B}_{\text{eff}}$ is the strength of the effective field $B_{\text{eff}}$ in Tesla (T). Our numerical estimation finds that
\begin{equation}\label{eq:mtot}
M_{\text{tot}} = 2.4\,\mathsf{B}_{\text{eff}} \times 10^{-4}\mu_B/V_{\text{u.c.}},
\end{equation}
where $\mu_B/V_{\text{u.c.}}$ denotes Bohr magneton per volume of the four-layer unit cell $V_{\text{u.c.}}$, with $V_{\text{u.c.}} = 2V_{\text{2D unit cell}}$. This provides an estimate of the strength of the weak FM order: for the magnetization in the TRSB phase to be non-detectable by a magnetometry device of sensitivity $10^{-10}$ T \cite{website}, the order parameter cannot exceed $B_{\text{eff}} =  6\times 10^{-5}$ T.

\section{pairing channels: Intralayer versus interlayer}\label{sec:two_pairing_channels}

\begin{figure}
\centering
\includegraphics[width=0.48\textwidth]{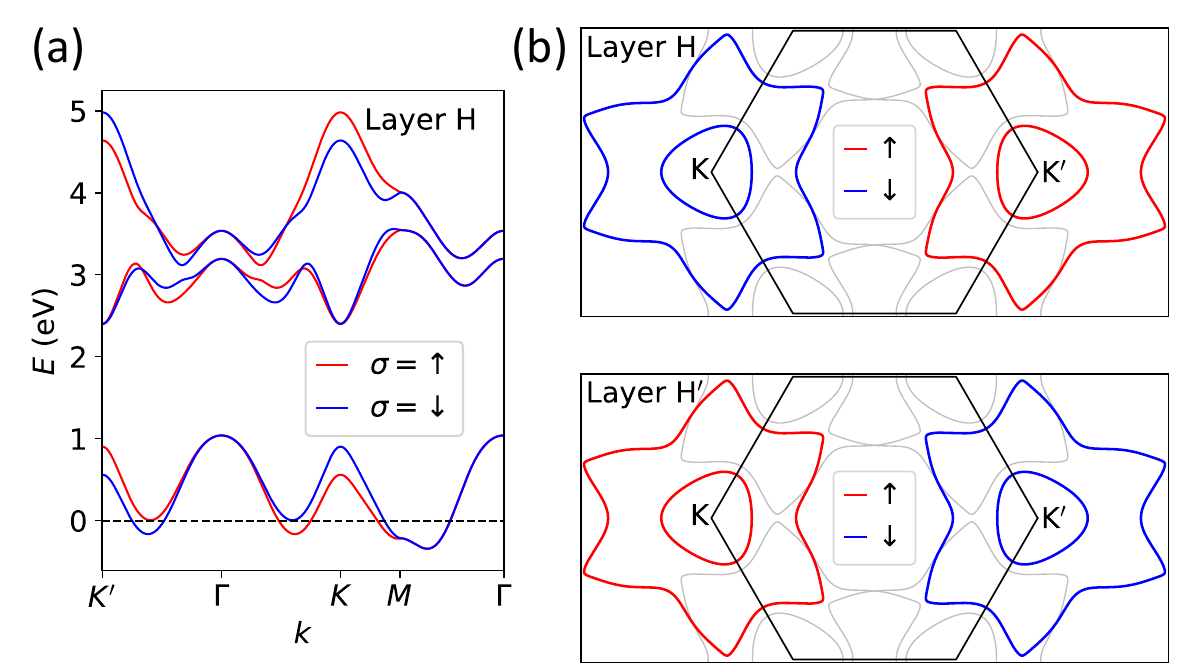}
\caption{Band structure of 4Hb-TaS$_2$. (a) Energy dispersion along high symmetry path in the Brillouin zone (BZ). (b) Fermi surfaces (FSs) for the $\mathrm{H}$ layer (upper panel) and the $\mathrm{H'}$ layer (lower panel). There are a smaller, near circular, hole-like FS and a larger, hexagonal, electron-like FS, concentered at the corners of the BZs, $\mathrm{K}$ and $\mathrm{K}'$. The complete set of FSs consists all those in red, blue and gray.}\label{fig:band}
\end{figure}

As the resistivity near the transition temperature exhibits BCS behavior with no substantial fluctuation regime \cite{doi:10.1126/sciadv.aax9480}, we here seek a pairing state consistent with a weak coupling BCS instability. This pairing is highly constrained by the Fermi surface (FS) geometry and symmetry. To see this, it is convenient to treat the layer index ($\mathrm{H},\mathrm{H'}$) as a good quantum number.
 The $(\uparrow,\mathrm{H})$ sector has a smaller, near circular, hole-like FS and a larger, hexagonal, electron-like FS, concentered at the $\mathrm{K}$ point of the $\mathrm{H}$-layer BZ. The $(\downarrow,\mathrm{H'})$ sector has similar FSs centered at $\mathrm{K}$ of the $\mathrm{H'}$-layer BZ, as shown in Fig.~\ref{fig:band}(b). The FSs from the two sectors exactly coincide with each other. Similarly, the other two sectors, $(\uparrow,\mathrm{H}')$ and $(\downarrow,\mathrm{H})$ have FSs centered at $\mathrm{K}'$. 

This FS geometry suggests that the only two possible pairing channels are
\begin{equation}\label{pairings}
\underbrace{(\uparrow,\mathrm{H},\bm{k})\leftrightarrow(\downarrow,\mathrm{H},-\bm{k})}_{\text{opposite spin, intralayer}}\text{ or }\, \underbrace{(\uparrow,\mathrm{H},\bm{k})\leftrightarrow(\uparrow,\mathrm{H}',-\bm{k})}_{\text{equal spin, interlayer}}.
\end{equation}
The former in Eq.~\eqref{pairings} describes the TRS-preserving, conventional BCS pairing. This channel is \emph{suppressed} by the breaking of TRS in the normal state: 
the chemical potentials are different for the up and down spins, $E_\uparrow - E_\downarrow \neq0$, and time-reversed electronic states are already energetically detuned, disfavoring the spin-singlet BCS pairing.
On the other hand, the latter in Eq.~\eqref{pairings} describes an inversion-symmetric, spin-triplet pairing, which remains a good pairing channel since inversion symmetry $i$ is unbroken in the normal state. This is a simple argument in favor of interlayer superconductivity in 4Hb-TaS$_2$.

Furthermore, we argue that fluctuations of the TRSB order parameter $\phi$ can mediate this pairing.
The effective action that couples the fluctuations of $\phi$ to the low-energy electrons is
\begin{equation}
S = \chi^{-1} \phi^2 + \phi(n_{\mathrm{H},\uparrow} - n_{\mathrm{H},\downarrow} + n_{\mathrm{H'},\uparrow} - n_{\mathrm{H'},\downarrow}),
\end{equation}
where $\chi>0$ is the susceptibility for $\phi$, and $n_{\ell,\sigma}$ denotes the electron number density for layer $\ell$ and spin $\sigma$. Upon integrating out $\phi$ one gets 
\begin{equation}
S_{\text{eff}}\sim - \chi (n_{\mathrm{H},\uparrow} - n_{\mathrm{H},\downarrow} + n_{\mathrm{H'},\uparrow} - n_{\mathrm{H'},\downarrow})^2.
\end{equation}
It contains the term $-\chi n_{\mathrm{H},\uparrow} n_{\mathrm{H'},\uparrow}$, which is attractive when decoupled in the spin-triplet intralayer pairing channel, and the term $\chi n_{\mathrm{H},\uparrow} n_{\mathrm{H},\downarrow}$, which is repulsive in the conventional BCS channel. Therefore the only possible pairing favored by the fluctuations of $\phi$ is the interlayer spin-triplet pairing.

\section{Symmetry of the Interlayer pairing}\label{Sec:Interlayer pairing}

\subsection{Summary of main points}

We now characterize the proposed interlayer spin-triplet pairing states using lattice symmetries. The point group symmetry $D_{6h}$ of the paramagnetic normal state is broken down to $C_{6h}$ by the FM order. The interlayer, equal-spin pairing is necessarily symmetric in the spin channel, and gives an antisymmetric orbital wave function living in the odd-parity irreps of $C_{6h}$. Furthermore, the pairing intrinsically involves different momentum layers in the 3D BZ: denote $\Delta_{\mathrm{H}\mathrm{H'}}$ ($\Delta_{\mathrm{H'}\mathrm{H}}$) as the pairing between $\mathrm{H}$ ($\mathrm{H'}$) and the adjacent $\mathrm{H'}$ ($\mathrm{H}$) layer above it, $\Delta_{\mathrm{H}\mathrm{H'}}$ and $\Delta_{\mathrm{H'}\mathrm{H}}$ are related by the twofold rotation $c_2$ in $C_{6h}$ (strictly speaking, the twofold screw). Depending on the eigenvalue of $c_2$ being $\pm1$, we have $\Delta_{\mathrm{H}\mathrm{H'}}=\pm\Delta_{\mathrm{H'}\mathrm{H}}$, and the gap has maximum amplitude (horizontal line node) on the $k_z=0$ or $k_z=\pi$ respectively ($k_z=\pi$ or $k_z=0$) plane in the 3D BZ. As we argue below, a single bilayer setup retains the essential feature of the interlayer pairing function obtained from a complete 3D pairing symmetry analysis. In this case $\Delta_{\mathrm{H}\mathrm{H'}}$ carries the irrep of $S_6$, generated by inversion and $c_3$, and is fully characterized by a function in the 2D momentum plane $\bm{k}=(k_x,k_y)$. The full 3D gap function can be easily recovered from $\Delta_{\mathrm{H}\mathrm{H'}}$ on the plane of maximum gap.

\subsection{Interlayer pairing in a bilayer system}

Here we discuss the symmetry classification of the gap function. Recall that normal state Hamiltonian $H_0$ preserves the full point group symmetry $D_{6h} = \langle c_3,m_x,m_z,i\rangle$. Let $\sigma^{0,1,2,3}$ denote the Pauli matrices for the spin space $\uparrow,\downarrow$ and $\mu^{0,1,2,3}$ denote those for the layer space $\mathrm{H}$ and $\mathrm{H'}$. The most general pairing is written as 
\begin{equation}
H_{\text{pairng}} = \sum_{\bm{k}} c^\dag_{\bm{k}} \Delta(\bm{k}) (c^\dag_{-\bm{k}})^T,
\end{equation}
with 
\begin{equation}
c^\dag_{\bm{k}} = (c^\dag_{\bm{k},\uparrow,\mathrm{H}},c^\dag_{\bm{k},\uparrow,\mathrm{H}'},c^\dag_{\bm{k},\downarrow,\mathrm{H}},c^\dag_{\bm{k},\downarrow,\mathrm{H}'}).
\end{equation}
Here $c^\dag_{\bm{k}}$ is the creation operator for the lowest electronic band where the Fermi energy lies, and
\begin{equation}\label{dstpa}
\Delta(\bm{k}) = \sum_{a=0,x,y,z}(\psi_a(\bm{k}) (i\sigma^y)+ \bm{d}_a (\bm{k}) \cdot \bm{\sigma}(i\sigma^y))\otimes \mu^a,
\end{equation}
$\psi_a(\bm{k})$ and $\bm{d}_a=(d_a^x(\bm{k}),d^y_a(\bm{k}),d^z_a(\bm{k}))$, $a=0,1,2,3$ are any complex functions of $\bm{k}$ to be constrained below.

Fermion anticommutation relation imposes the parity condition
$\Delta({\bm{k}}) = - \Delta^T(-{\bm{k}})$, or
\begin{equation}\label{oooree}
\begin{aligned}
\psi_0(\bm{k}) = \psi_0(-\bm{k}),\quad
\psi_1(\bm{k}) = \psi_1(-\bm{k}),\\
\psi_2(\bm{k}) = -\psi_2(-\bm{k}),\quad
\psi_3(\bm{k}) = \psi_3(-\bm{k}),\\
\bm{d}_0(\bm{k}) = -\bm{d}_0(-\bm{k}),\quad
\bm{d}_1(\bm{k}) = -\bm{d}_1(-\bm{k}),\\
\bm{d}_2(\bm{k}) = \bm{d}_2(-\bm{k}),\quad
\bm{d}_3(\bm{k}) = -\bm{d}_3(-\bm{k}).
\end{aligned}
\end{equation}
Since we will be interested in the interlayer, spin triplet channels $|\uparrow \mathrm{H}\,\rangle|\uparrow \mathrm{H'}\,\rangle$ and $|\downarrow \mathrm{H}\,\rangle|\downarrow \mathrm{H}'\,\rangle$, the relevant pairing functions to be considered are $d^x_1(\bm{k})$, $d^y_1(\bm{k})$, $d^x_2(\bm{k})$, and $d^y_2(\bm{k})$. Explicitly, the relevant pairing terms are (suppressing the momentum dependence of $d^{x,y}_{1,2}$)
\begin{subequations}\label{S34}
\begin{align}
\Delta_{\text{interlayer,}\uparrow}=&\sum_{\bm{k}}( -d^x_1+id^y_1+d^y_2+id^x_2)c^\dag_{\bm{k},\mathrm{H},\uparrow}c^\dag_{-\bm{k},\mathrm{H}',\uparrow}\notag\\
&+
\sum_{\bm{k}}( -d^x_1+id^y_1-d^y_2-id^x_2)c^\dag_{\bm{k},\mathrm{H'},\uparrow}c^\dag_{-\bm{k},\mathrm{H},\uparrow},\\
\Delta_{\text{interlayer,}\downarrow} =& \sum_{\bm{k}}(d^x_1+id^y_1+d^y_2-id^x_2) c^\dag_{\bm{k},\mathrm{H},\downarrow}c^\dag_{-\bm{k},\mathrm{H}',\downarrow}\notag\\
&+ \sum_{\bm{k}}(d^x_1+id^y_1-d^y_2+id^x_2) c^\dag_{\bm{k},\mathrm{H'},\downarrow}c^\dag_{-\bm{k},\mathrm{H},\downarrow}.
\end{align}
\end{subequations}
Treating layer index as another pseudospin index \footnote{We will comment on the validity of treating the layer index as a pseudospin index in subsection \ref{sec:3Dtreatment}.}, we see that $d^{x,y}_1(\bm{k})$ produces the spin triplet, layer triplet pairing, while $d^{x,y}_2(\bm{k})$ produces the spin triplet, layer singlet pairing. This implies that $\bm{d}_1(\bm{k})$ is an odd function of momentum in the spin-layer space while $\bm{d}_2(\bm{k})$ is an even function momentum in the spin-layer space.

Now let us examine how $\bm{d}_{1,2}$ transform under elements of the point group $D_{6h}$. Write
\begin{subequations}\label{dvectors12}
\begin{align}
\bm{d}_1 &= d_1^x(\bm{k})\bm{\hat{x}} + d_1^y(\bm{k})\bm{\hat{y}},\\
\bm{d}_2 &= d_2^x(\bm{k}) \bm{\hat{\alpha}}+ d_2^y(\bm{k})\bm{\hat{\beta}},
\end{align}
\end{subequations}
where the basis function $\bm{\hat{x}}$ ($\bm{\hat{y}}$) denotes the layer-triplet, spin-triplet (layer-triplet, spin-singlet) wave function, while $\bm{\hat{\alpha}}$ ($\bm{\hat{\beta}}$) denotes the layer-singlet, spin-triplet (layer-triplet, spin-singlet) wave function. We have turned off the $d^z$ component as it gives the spin-mixing channel $|\uparrow \downarrow\rangle +|\downarrow \uparrow\rangle$ that is irrelevant for us purposes. The following transformation rule can derived from that of the electron creation operators:
\begin{subequations}\label{transic33}
\begin{align}
i&\colon ~~~\bm{d}_1(\bm{k})\rightarrow \bm{d}_1(-\bm{k}),\quad \bm{d}_2(\bm{k})\rightarrow -\bm{d}_2(-\bm{k}),\\
c_3&\colon ~~~\bm{d}_{1,2}(\bm{k}) \rightarrow \mathbf{R}_{\text{2D}}\bm{d}_{1,2}(c_3^{-1}(\bm{k})),
\end{align}
\end{subequations}
where $\mathbf{R}_{\text{2D}} = \left(\begin{smallmatrix}-\frac{1}{2}&-\frac{\sqrt{3}}{2}\\ \frac{\sqrt{3}}{2} & - \frac{1}{2}\end{smallmatrix}\right)$ is the usual SO(2) rotation matrix for $c_3$. The transformation under $m_x$ and $m_z$ can be obtained in a similar manner. Importantly, we see that
\begin{equation}
i\colon ~~~\bm{\hat{x}},\bm{\hat{y}}\rightarrow \bm{\hat{x}},\bm{\hat{y}},\quad \bm{\alpha},\bm{\hat{\beta}}\rightarrow -\bm{\hat{\alpha}},-\bm{\hat{\beta}},
\end{equation}
in other words, the basis vectors $\bm{\hat{\alpha}},\bm{\hat{\beta}}$ transform as (polar) vectors, while the basis vectors $\bm{\hat{x}},\bm{\hat{y}}$ transform as pseudovectors. These rules allow us to construct the basis functions for the pairing functions $\bm{d}_{1,2}$.

\subsection{Interlayer pairing in the 3D bulk}\label{sec:3Dtreatment}

The bulk 4Hb-TaS$_2$ contains multiple layers and a 3D treatment of the pairing functions is warranted. Below we outline this analysis.  To start with, we absorb the layer index into momentum. We temporarily suppress spin indices for simplicity; the spin indices can be easily restored (see below). We define a new version of Fourier transform that includes both the $\mathrm{H}$ and $\mathrm{H'}$ layers as
\begin{subequations}\label{Fourierhhp}
\begin{align}
c^\dag_{\bm{r},\ell,\mathrm{H}} &= \frac{1}{\sqrt{2N}}\sum_{\bm{k},k_z} e^{-i(\bm{k}\cdot \bm{r}+k_z 2\ell c_{\frac{1}{2}})}c^\dag_{\bm{k},k_z},\\
c^\dag_{\bm{r},\ell,\mathrm{H'}} &= \frac{1}{\sqrt{2N}} \sum_{\bm{k},k_z} e^{-i(\bm{k}\cdot \bm{r}+k_z (2\ell+1)c_{\frac{1}{2}}}c^\dag_{\bm{k},k_z},
\end{align}
\end{subequations}
where $\bm{k}=(k_x,k_y)$ is the 2D momentum, and $k_z \in [-\frac{\pi}{c_{\frac{1}{2}}},\frac{\pi}{c_{\frac{1}{2}}}]$, where $c=2c_{\frac{1}{2}}$ is the distance between two $\mathrm{H}$ layers, while $c_{\frac{1}{2}}$ is the distance between adjacent $\mathrm{H}$ and $\mathrm{H'}$ layers. Under this definition, the in-plane gap function $\Delta_{12}(\bm{k})$ and $\Delta_{21}(\bm{k})$ acquires $k_z$ dependence
\begin{equation}\label{ftpa}
\begin{aligned}
&\sum_\ell \Delta_{12}c^\dag_{\bm{k},\ell,\mathrm{H}}c^\dag_{-\bm{k},\ell,\mathrm{H'}}+\Delta_{21}c^\dag_{\bm{k},\ell,\mathrm{H'}}c^\dag_{-\bm{k},\ell+1,\mathrm{H}}+h.c.\\
&=\sum_{k_z} (\Delta_{12} e^{-ik_z c_{\frac{1}{2}}} +\Delta_{21} e^{ik_z c_{\frac{1}{2}}}) c^\dag_{\bm{k},k_z}c^\dag_{-\bm{k},-k_z}+h.c.,
\end{aligned}
\end{equation}
Let us now examine how the gap functions transform under the layer pseudospin index. According to Eqs.~\eqref{Fourierhhp}, interchanging layer indices $\mathrm{H}\leftrightarrow \mathrm{H'}$ amounts to multiplying by a phase $c^\dag_{\bm{k},k_z}\rightarrow e^{ik_z c_{\frac{1}{2}}}c^\dag_{\bm{k},k_z}$. This phase multiplication obviously leaves Eq.~\eqref{ftpa} unchanged, meaning that absorbing the layer indices into momentum will only retain the channels that are symmetric in the layer pseudospins (i.e. layer pseudospin triplet), in contrast to the more general treatment in Eq.~\eqref{dstpa}. The retaining of only symmetric channels in the layer indices makes sense, since the layer indices are spatially locked with momentum and is strictly speaking not an independent internal degree of freedom (had the two layers sit on top each other without any spatial displacement the layer index would have been a genuine free index).

Now, we further define
\begin{subequations}
\begin{align}
d_1(\bm{k},k_z) &= (\Delta_{12}(\bm{k})+\Delta_{21}(\bm{k}))  \cos k_z c_{\frac{1}{2}},\\
d_2(\bm{k},k_z) &= i(-\Delta_{12}(\bm{k})+\Delta_{21}(\bm{k}))\sin k_z c_{\frac{1}{2}},
\end{align}
\end{subequations}
upon restoring spin indices $d\rightarrow (\psi,\bm{d})$, the above $d_1\rightarrow \bm{d}_1$ and $d_2\rightarrow \bm{d}_2$ correspond to the vectors $\bm{d}_{1,2}$ in Eq.~\eqref{dvectors12}. Fermion anticommutation relation requires that 
\begin{equation}
\begin{aligned}
&\Delta_{12}(\bm{k},k_z)e^{-ik_zc_{\frac{1}{2}}}+\Delta_{21}(\bm{k},k_z)e^{ik_zc_{\frac{1}{2}}}\\
&=-\Delta_{12}(-\bm{k},-k_z)e^{ik_zc_{\frac{1}{2}}}+\Delta_{21}(-\bm{k},-k_z)e^{-ik_zc_{\frac{1}{2}}}.
\end{aligned}
\end{equation}
 If $\Delta_{12}$ and $\Delta_{21}$ do not depend on $k_z$, then we must have $\Delta_{21}(\bm{k}) = -\Delta_{12}(-\bm{k})$.

Under the twofold screw $s_2$ (denoted as $c_2$ in the point group notation) we have
\begin{equation}
c_2\colon ~~~d_1(\bm{k},k_z)\rightarrow d_1(-\bm{k},k_z),\quad
d_2(\bm{k},k_z)\rightarrow - d_2(-\bm{k},k_z),
\end{equation}
this means that $d_1$ and $d_2$ transform under the even and odd irreps of $c_2$, respectively. 

The superconducting state results from a ferromagnetically ordered state. Such a state has point group symmetry $C_{6h}$, generated by $c_3$, $i$, and $c_2$.  The lattice basis functions corresponding to different irreps of $C_{6h}$ are given in Table \ref{tab:basis_function}

\begin{table*}[!thb]
\caption{Table for the lowest order basis functions for irreps of $C_{6h}= \langle c_3,i,c_2\rangle=\langle c_3,i,m_z\rangle$. The representative elements in the table are $c_3$ and $c_2 = i\circ m_z$. We defined $\bm{\hat{x}}_\pm = \bm{\hat{x}}\pm i \bm{\hat{y}}$, $K_+ = \sum_j \omega^j \sin \bm{k}\cdot \bm{e}_j$, $K_-=\sum_j (\omega^j)^* \sin \bm{k}\cdot \bm{e}_j$, $K^2_+= \sum_j (\omega^j)^* \cos \bm{k}\cdot \bm{e}_j$, $K^2_-= \sum_j \omega^j \cos \bm{k}\cdot \bm{e}_j$, where $j=0,1,2$, $\omega = e^{i\frac{2\pi}{3}}$, $\bm{e}_0 = (1,0)$, $\bm{e}_1=(-\frac{1}{2},\frac{\sqrt{3}}{2})$, and $\bm{e}_2=(-\frac{1}{2},-\frac{\sqrt{3}}{2})$. Note at small momentum, $K_\pm \sim k_\pm \equiv k_x \pm i k_y$, and $K^2_\pm \sim k^2_\pm$. The vertical lattice constant is set to unity.}
\label{tab:basis_function}
\begin{tabular}{c|cc|c|cc}
\hline
$C_{6h}$&\multicolumn{2}{c|}{Character}&Basis function for interlayer $|\uparrow\uparrow\rangle$ channel&Basis function for interlayer $|\downarrow\downarrow\rangle$ channel\\
\hline
Irrep& $c_3$ & $c_2$ &  $(d^x-id^y)\bm{\hat{x}}_+$ &   $(d^x+i d^y)\bm{\hat{x}}_-$  \\ 
\hline
$A_{\text{g}}$& $+1$ & $+1$ &$K_+^2\cos \frac{k_z}{2}\bm{\hat{x}}_+$ & $K_-^2 \cos \frac{k_z}{2}\bm{\hat{x}}_-$\\
$B_{\text{g}}$& $+1$ & $-1$ &$K_-\sin \frac{k_z}{2}\bm{\hat{x}}_+$&$K_+\sin \frac{k_z}{2}\bm{\hat{x}}_-$\\
$E_{\text{1g}}$& $-1$ & $-2$ &$K_+\sin \frac{k_z}{2}\bm{\hat{x}}_+,K_-^3\sin \frac{k_z}{2}\bm{\hat{x}}_+$&$K_-\sin\frac{k_z}{2}\bm{\hat{x}}_-,K^3_+\sin \frac{k_z}{2}\bm{\hat{x}}_-$\\
$E_{\text{2g}}$& $-1$ & $+2$ &$\cos \frac{k_z}{2}\bm{\hat{x}}_+,K_-^2\cos \frac{k_z}{2}\bm{\hat{x}}_+$&$\cos\frac{k_z}{2}\bm{\hat{x}}_-,K_+^2\cos \frac{k_z}{2}\bm{\hat{x}}_-$\\
\hline
$A_{\text{u}}$& $+1$ & $+1$ &$K_-\cos \frac{k_z}{2}\bm{\hat{x}}_+$&$K_+ \cos \frac{k_z}{2}\bm{\hat{x}}_-$\\
$B_{\text{u}}$& $+1$ & $-1$ &$K_+^2\sin \frac{k_z}{2}\bm{\hat{x}}_+$&$K_-^2 \sin \frac{k_z}{2}\bm{\hat{x}}_- $\\
$E_{\text{1u}}$& $-1$ & $-2$  &$\sin \frac{k_z}{2}\bm{\hat{x}}_+,K^2_-\sin \frac{k_z}{2}\bm{\hat{x}}_+$&$\sin\frac{k_z}{2}\bm{\hat{x}}_-,K_+^2\sin \frac{k_z}{2}\bm{\hat{x}}_-$\\
$E_{\text{2u}}$& $-1$ & $+2$  &$K_+\cos \frac{k_z}{2}\bm{\hat{x}}_+,K_-^3\cos \frac{k_z}{2}\bm{\hat{x}}_+$&$K_-\cos \frac{k_z}{2}\bm{\hat{x}}_-,K_+^3\cos \frac{k_z}{2}\bm{\hat{x}}_-$\\
\hline
\end{tabular}
\end{table*}

The above symmetry analysis classifies the gap function $\bm{d}$ but is not valid for the pairing function $\Delta_{12}(\bm{k})$, because the operation $c_2$ transforms $\Delta_{12}(\bm{k})$ to $\Delta_{21}(-\bm{k})$. In fact, a single $\Delta_{12}(\bm{k})$ transforms to itself only under $c_3$ and $i$, hence $\Delta_{12}(\bm{k})$ itself is classified by the symmetry group $S_6$ generated by $c_3$ and $i$. The irreps of $\bm{d}$ can be easily recovered from those of $\Delta_{12}(\bm{k})$ by further specifying the character of $c_2$ (being $\pm 1$, which further specifies the dominant $k_z$ momentum plane on which the gap function amplitude is maximal). The classification of basis functions for $\Delta_{12}$ is given in Table \ref{tab:s6basis_function}.

\begin{table*}
\caption{Table for the lowest order basis functions for irreps of $S_6= \langle c_3,i\rangle$. We defined $k_\pm \equiv k_x \pm i k_y$. The vertical lattice constant is set to unity.}
\label{tab:s6basis_function}
\begin{tabular}{c|c|c|cc}
\hline
$S_6$&Character&Basis function for interlayer $|\uparrow\uparrow\rangle$ channel&Basis function for interlayer $|\downarrow\downarrow\rangle$ channel\\
\hline
Irrep& $c_3$ &  $\Delta_{12}(\bm{k},k_z)\bm{\hat{x}}_+$ &   $\Delta_{12}(\bm{k},k_z)\bm{\hat{x}}_-$  \\ 
\hline
$A_{\text{g}}$& $+1$ &$K_+^2\cos \frac{k_z}{2}\bm{\hat{x}}_+$, $K_-\sin \frac{k_z}{2}\bm{\hat{x}}_+$ & $K_-^2 \cos \frac{k_z}{2}\bm{\hat{x}}_-,K_+\sin\frac{k_z}{2}\bm{\hat{x}}_-$\\
$E_{\text{g}}$& $-1$ &$K_+\sin \frac{k_z}{2}\bm{\hat{x}}_+,K_-^3\sin \frac{k_z}{2}\bm{\hat{x}}_+, K^2_-\cos \frac{k_z}{2}\bm{\hat{x}}_+,\cos\frac{k_z}{2}\bm{\hat{x}}_+$&$K_-\sin\frac{k_z}{2}\bm{\hat{x}}_-,K^3_+\sin \frac{k_z}{2}\bm{\hat{x}}_-,K^2_+\cos\frac{k_z}{2}\bm{\hat{x}}_-,\cos\frac{k_z}{2}\bm{\hat{x}}_-$\\
\hline
$A_{\text{u}}$& $+1$ &$K_-\cos \frac{k_z}{2}\bm{\hat{x}}_+$, $K_+^2\sin \frac{k_z}{2}\bm{\hat{x}}_+$&$K_+ \cos \frac{k_z}{2}\bm{\hat{x}}_-,K_-^2\sin\frac{k_z}{2}\bm{\hat{x}}_-$\\
$E_{\text{u}}$& $-1$ &$\sin \frac{k_z}{2}\bm{\hat{x}}_+,K^2_-\sin \frac{k_z}{2}\bm{\hat{x}}_+,K_+^3\cos \frac{k_z}{2}\bm{\hat{x}}_+,K_+\cos\frac{k_z}{2}$&$\sin\frac{k_z}{2}\bm{\hat{x}}_-,K_+^2\sin \frac{k_z}{2}\bm{\hat{x}}_-,K^3_-\cos\frac{k_z}{2}\bm{\hat{x}}_-,K_-\cos\frac{k_z}{2}$\\
\hline
\end{tabular}
\end{table*}

\subsection{Relation between a 2D gap function and a 3D gap function}

The above analysis shows that the gap function $\Delta_{12}(\bm{k},k_z)$ is intrinsically 3D and is dominant on the $k_z=0$ or $k_z=\pi$ layers. While a complete analysis of superconductivity should involve the full 3D BZ, below we justify the calculation in a 2D BZ layer. This will allow us to study the magnetization in the superconducting state with an approximate 2D gap function $\Delta_{\bm{k},\uparrow}$ on a 2D BZ in the next sections. 

The gap function $\Delta_{12}$ always lives in the odd parity representations. As Table \ref{tab:s6basis_function} suggests, the winding of a single Fermi surface is enough to specify whether it lives on the $k_z=0$ or the $k_z=\pi$ plane in the 3D BZ. For example, if gap winding on the inner and outer Fermi surfaces differ by multiples of $12\pi$, then both lives on the same $k_z$ plane in the 3D BZ; otherwise, the gap winding on the inner and outer Fermi surfaces must differ by a multiples of $\pm 3\pi$, $\pm 9 \pi$ and so on, and one of the will live in the $k_z=0$ plane while the other on the $k_z=\pi$ plane. However, as the orbital magnetization receives contributed mainly from the vicinity of the FSs, we can ``superpose'' the $k_z=0$ and $k_z=\pi$ planes to get a single 2D BZ, and the calculation of magnetization on this plane should match that of a full 3D calculation. For this reason, we have been using a 2D model in the main text and the sections above, treating the layer indices $\mathrm{H}$ and $\mathrm{H'}$ as internal indices. The calculated magnetization should be understood as the result for a full 3D magnetization (i.e. averaged over the momentum planes indexed by $k_z$).

As a side comment, when the gap function lives on $k_z=\pi$ plane, the $k_z=0$ plane remains metallic with a nodal line Fermi surface, yet this nodal line structure may be gapped out by an interlayer tunneling and an intralayer pairing.

\section{Physical consequences of the interlayer pairing}\label{sec:consequence_topology_unpair}

\subsection{Topology}

Denote the pairing gap function between $\mathrm{H}$ and the adjacent $\mathrm{H'}$ layer above it as
\begin{equation}
\Delta_{\bm{k},\uparrow} = \langle c^\dag_{\bm{k},\mathrm{H},\uparrow} c^\dag_{-\bm{k},\mathrm{H'},\uparrow}\rangle.
\end{equation}
Due to the breaking of TRS,  all the irreps of $S_6$ become one-dimensional and allow chiral ansatze in the equal-spin-up channels.  
Due to the discrete rotation $c_3$, the orbital angular momentum carried by $\Delta_{\bm{k},\uparrow}$ is defined modulo $3$;
Equivalently, the gap windings on the inner and outer FSs centered at $\mathrm{K}$ can only differ by multiples of $6\pi$.
Furthermore, this winding difference, or the \emph{total gap winding} upon including the sign $+$/$-$ for electron-/hole-like FSs, exactly defines the (gauge invariant) Chern number $c_{\text{BdG}}$ for the Bogoliubov-de Gennes (BdG) bands. From this we conclude that \emph{the BdG Chern number can only be multiples of $3$.}

To illustrate the above claims, we perform a microscopic calculation by solving the BCS mean-field equation for the interlayer pairing gap function (see Appendix \ref{app:gap_free_energy} for more detail)
\begin{figure}
\centering
\includegraphics[width=0.43\textwidth]{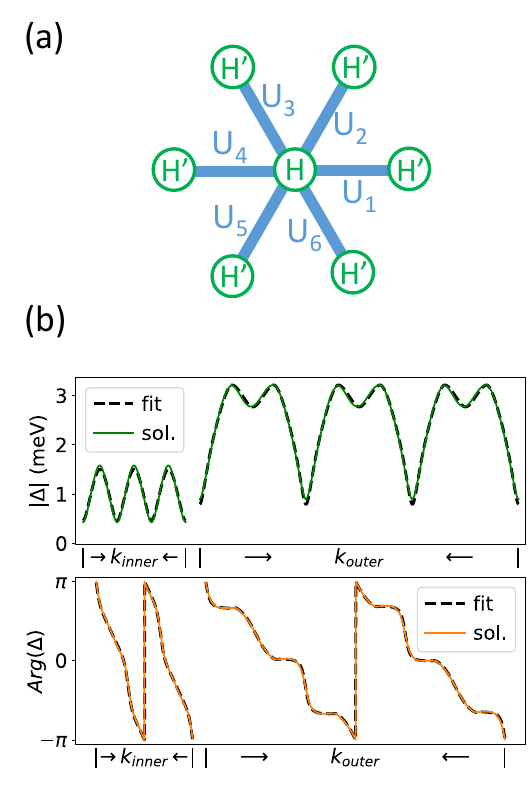}
\caption{
(a) Illustration of the real space, NN interlayer pairing between layers $\mathrm{H}$ and $\mathrm{H'}$. (b) The gap function $\Delta$ for the $c_{\text{BdG}}=0$ ansatz. The upper (lower) panel shows the amplitude (phase) of $\Delta$ on both the inner and outer FSs. The solid lines are solved from Eq.~\eqref{gapfunc} and the dashed lines are obtained from a NN interlayer pairing ansatz as a lattice approximation. The gap minimum is set to the experimental value $0.44$ meV \cite{nayak2021evidence}.
}\label{fig:gap}
\end{figure}
\begin{equation}\label{gapfunc}
\Delta_{\bm{k},\uparrow}
=-\int\frac{d^2k'}{(2\pi)^2}
V_{\bm{k}-\bm{k}'}\langle u_{\bm{k},\uparrow,\mathrm{H}}|u_{\bm{k}',\uparrow,\mathrm{H}}\rangle^{2} \frac{\tanh\frac{\varepsilon_{\bm{k}'}}{2T}}{2 \varepsilon_{\bm{k}'}}\Delta_{\bm{k}',\uparrow},
\end{equation}
here $|u_{\bm{k},\uparrow,\mathrm{H}}\rangle$ is the normal state Bloch wave function for spin $\uparrow$ and layer $\mathrm{H}$, $\varepsilon_{\bm{k}}$ is the dispersion of the Bogoliubov quasi-particles, and we assumed for simplicity a constant attractive interaction $V_{\bm{k}}\!=\!V\!<\!0$.
The solution is shown in Fig.~\ref{fig:gap}(b). This ansatz lives in the $E_{u}$ irrep of $S_6$ and has a $-4\pi$ winding on both the inner and outer FSs, but has a vanishing Chern number, $c_{\text{BdG}}\!=\!0$.

\subsection{Unpaired minority component} 

A direct consequence of the interlayer spin-polarized pairing is that the other spin component remains unpaired and forms a gapless FS below $T_c$.
This can be understood from a Landau free energy analysis. Denote the real space, coarse grained interlayer spin-polarized pairing order parameters as $\Delta_\sigma$. At quadratic level the allowed terms are 
\begin{equation}
F_2 = \sum_{\sigma} r_\sigma |\Delta_\sigma|^2,
\end{equation}
where $r_\uparrow$ and $r_\downarrow$ are generally different because TRS is already broken in the normal state. As a consequence the critical temperatures for pairing of the two spin components $T_{c\uparrow}$ and $T_{c\downarrow}$ are also different. Note that there is no proximity coupling between the paired and unpaired components because the cross-term $\Delta_\uparrow \Delta_\downarrow^*$ is forbidden due to spin-$S_z$ conservation.
Therefore the unpaired spin component remains gapless between the upper and lower critical temperatures.

The prediction of two transition temperatures raises a possible discrepancy between this theory and experimental observations.  Since the breaking of TRS in the normal state is assumed to be weak, one might expect it would lead to only a small difference between the two critical temperatures. However, experiments have not observed the lower transition temperature down to $0.5$ K \cite{doi:10.1126/sciadv.aax9480}\footnote{It is worth noting that there may be an anomalously wide separation between the two transition temperatures even at the quadratic level because the bands are close to a Van Hove singularity. In a weakly ferromagnetic state the majority component gets closer to it, while the minority gets further, which can lead to a sizable difference even for small magnetization.}. 
The discrepancy may be resolved by considering  the quartic terms of the Landau energy: 
\begin{equation}\label{eq:f4}
F_4=u(|\Delta_\uparrow|^4+|\Delta_\downarrow|^4) + v|\Delta_\uparrow|^2|\Delta_\downarrow|^2.
\end{equation}
While $u$ is a property of the electronic band structure, the other coefficient $v\propto \chi^2$ is proportional to the square of the chiral susceptibility $\chi$ (see Appendix \ref{app:gap_free_energy} for derivation). When the fluctuations of $\phi$ are sufficiently strong such that $v \!>\!2u\!>\!0$, a total suppression of the minority spin pairing can be achieved.

\section{Magnetization in the superconducting state}\label{sec:mag}
 
Having discussed general features of the interlayer pairing states we now come to the central discussion of how and to what extent such pairing amplifies the normal state magnetization.
To this end, we derive in Appendix \ref{app:mag_form_SC} expressions for the magnetization in the interlayer pairing state. Importantly, the orbital magnetization explicitly contains a Berry curvature contribution, which traces back to the orbital angular momentum of the pairing state (A similar relation is known for the normal state \cite{PhysRevLett.99.197202}). One consequence is that a nonzero $c_{\text{BdG}}$ \emph{enhances} orbital magnetization, as we will show in detail below.

\subsection{Magnetization in the interlayer pairing state}

The bulk magnetization consists of three parts:
\begin{equation}\label{MM123}
M^z = M^z_{\text{orb,t-b}} + M^z_{\text{orb,atom}} + M^z_{\text{spin}},
\end{equation}
where the last two terms are the atomic angular momentum and atomic spin contribution to the magnetization, which can be unambiguously written as
\begin{widetext}
\begin{equation}
M^z_{\text{orb,atom}}
=
-\frac{\mu_B}{V_{\text{2D unit cell}}}  \mathrm{ReTr}\left[
\int_{\text{BZ}} \frac{d^2k}{(2\pi)^2}
U^\dag_{\text{BdG}}\begin{pmatrix}L^z&0\\0&-L^z\end{pmatrix}\begin{pmatrix} {[U_{\text{BdG}}]}_{1:3} f_{\bm{k}} \\{[U_{\text{BdG}}]}_{4:6} (1-f_{\bm{k}}) \end{pmatrix}\right],
\end{equation}
\begin{equation}
M^z_{\text{spin}}
=\frac{g_s}{2} 
\frac{\mu_B}{V_{\text{2D unit cell}}} \mathrm{ReTr}\left[
\int_{\text{BZ}} \frac{d^2k}{(2\pi)^2}
U^\dag_{\text{BdG}} \begin{pmatrix}{[U_{\text{BdG}}]}_{1:3} f_{\bm{k}} \\{[U_{\text{BdG}}]}_{4:6} (1-f_{\bm{k}}) \end{pmatrix}\right],
\end{equation}
\end{widetext}
where $U_{\text{BdG}}$ diagonalizes the Hamiltonian
\begin{equation}\label{hbdgdiag}
\mathcal{H}_{\text{BdG}}(\bm{k}) = U_{\text{BdG}}(\bm{k}) \mathcal{E}_{\bm{k}} U^\dag_{\text{BdG}}(\bm{k}),
\end{equation}
$f_{\bm{k}} = f(\mathcal{E}_{\bm{k}})$ is Fermi-Dirac distribution (now a diagonal matrix), and $[U_{\text{BdG}}]_{1:3}$ denotes the first three rows of $U_{\text{BdG}}$.

The first term in Eq.~\eqref{MM123}, $M^z_{\text{orb,t-b}}$, denotes the orbital magnetization due to hopping and pairing. It has the form
\begin{widetext}
\begin{equation}
M^z_{\text{orb,t-b}}
=\frac{e}{\hbar}\mathrm{ImTr}
\left[\int_{\text{BZ}} \frac{d^2k}{(2\pi)^2}
\partial_{\bm{k}}
U^\dag_{\text{BdG}} \times \left(\mathcal{E}1_{6\times 6} + \mathcal{H}_{\text{BdG}}\Big|_{\Delta=0}\right)\begin{pmatrix} \partial_{\bm{k}}[U_{\text{BdG}}]_{1:3,:} f_{:,\bm{k}} \\\partial_{\bm{k}}[U_{\text{BdG}}]_{4:6,:} (1-f_{:,\bm{k}}) \end{pmatrix}\right].
\end{equation}
\end{widetext}

\subsection{Candidate pairing ansatze}

Now we analyze the pairing in real space. We assume equal spin, interlayer pairing, i.e. the pairing is only between the same spin species of the $\mathrm{H}$ layer and the $\mathrm{H}'$ layer. For the rest of this section, we will focus on the spin up sector and quite often we will omit the spin index $\sigma=\uparrow$.
\begin{widetext}
\begin{equation}\label{pairhe}
H_{\text{pairing},\uparrow}
=
\sum_{\bm{r},\bm{\delta}}
(d^\dag_{z^2,\uparrow,\bm{r},\mathrm{H}},d^\dag_{xy,\uparrow,\bm{r},\mathrm{H}},d^\dag_{x^2-y^2,\uparrow,\bm{r},\mathrm{H}})\Delta_{\bm{\delta}}
\left(\begin{array}{c} d^\dag_{z^2,\uparrow,\bm{r}+\bm{\delta},\mathrm{H}'}\\d^\dag_{xy,\uparrow,\bm{r}+\bm{\delta},\mathrm{H}'}\\d^\dag_{x^2-y^2,\uparrow,\bm{r}+\bm{\delta},\mathrm{H}'}\end{array}\right)+h.c.,
\end{equation}
\end{widetext}
where $\Delta_{\bm{\delta}}$ is a $3\times 3$ matrix denoting the $\bm{\delta}$-neighbor pairing in the orbital space. We model $\Delta_{\bm{\delta}}$ phenomenologically by the following Ansatz
\begin{equation}
\Delta_{\text{vert}} = 0_{3\times 3},~~~
\Delta_{U_1}
=[\Delta_1]_{i,j=1}^3,~~~
 \Delta_{U_4}
=[\Delta_2]_{i,j=1}^3,
\end{equation}
and the other four nearest-neighbor pairing matrices are related to $\Delta_{U_1}$ and $\Delta_{U_4}$ by (see Fig.~\ref{fig:gap} for the definition of the bonds $U_i$)
\begin{equation}
\begin{aligned}
\Delta_{U_6} =  e^{i\frac{2\pi}{3}} \mathbf{R}^T \Delta_{U_4}\mathbf{R},\quad 
\Delta_{U_2} = e^{-i\frac{2\pi}{3}} \mathbf{R} \Delta_{U_4}\mathbf{R}^T,\\
\Delta_{U_3} = e^{i\frac{2\pi}{3}} \mathbf{R}^T \Delta_{U_1} \mathbf{R},\quad 
\Delta_{U_5}= e^{-i\frac{2\pi}{3}} \mathbf{R} \Delta_{U_1} \mathbf{R}^T.
\end{aligned}
\end{equation}

The BdG Hamiltonian reads 
\begin{equation}\label{bdgbd}
\begin{aligned}
H_{\text{BdG},\uparrow} &= H_{\mathrm{H},\uparrow} + H_{\mathrm{H'},\uparrow}+
H_{\text{pairing},\uparrow} \\
&= \sum_{\bm{k}}(d^\dag_{\bm{k},\uparrow,\mathrm{H}}, d^T_{-\bm{k},\uparrow,\mathrm{H}'})\mathcal{H}_{\text{BdG}}(\bm{k})\begin{pmatrix}{c} d_{\bm{k},\uparrow,\mathrm{H}}\\ \big(d^\dag_{-\bm{k},\uparrow,\mathrm{H}'}\big)^T\end{pmatrix}
\end{aligned}
\end{equation}
where we defined the Hamiltonian matrix \begin{equation}\label{bdg1stq}
\mathcal{H}_{\text{BdG}}(\bm{k}) = \left(\begin{array}{cc}
\mathcal{H}_{\mathrm{H}}(\bm{k}) & \Delta^{\text{orb}}(\bm{k}) \\ \left(\Delta^{\text{orb}}(\bm{k})\right)^\dag & -\mathcal{H}^T_{\mathrm{H}}(\bm{k})\end{array}\right)
\end{equation}
with the help of Eq.~\eqref{relationbetweenlayer}. $\Delta^{\text{orb}}({\bm{k}})$ is the pairing matrix in momentum space which  we keep only up to nearest-neighbor bonds and has the form
\begin{equation}
\Delta^{\text{orb}}({\bm{k}}) =
\Delta_{\text{vert}}+
\sum_{i=1}^6 \Delta_{U_i} e^{i\bm{k}\cdot \bm{\delta}_i},
\end{equation}
where $\bm{\delta}_i$ are the six nearest-neighbor bonds.

We project $\Delta^{\text{orb}}_{\bm{k}}$ to the lowest band to obtain the gap function of that band, $\Delta(\bm{k})$:
\begin{equation}\label{fitdd}
\Delta(\bm{k})=   u^\dag_{\bm{k},\mathrm{H}} \Delta^{\text{orb}}(\bm{k}) u^*_{-\bm{k},\mathrm{H}'} = u^\dag_{\bm{k},\mathrm{H}} \Delta^{\text{orb}}(\bm{k}) u^*_{\bm{k},\mathrm{H}},
\end{equation}
where we have made use of the inversion symmetry for the normal state wave function $u_{\bm{k},\mathrm{H}} = u_{-\bm{k},\mathrm{H}'}$. The relation \eqref{fitdd} established the connection between our effective theory and the microscopic orbital pairing in the lattice.  The symmetry properties of the BdG Hamiltonian will be discussed in the next section.

\begin{figure*}[!thb]
\centering
\includegraphics[width=0.9\linewidth]{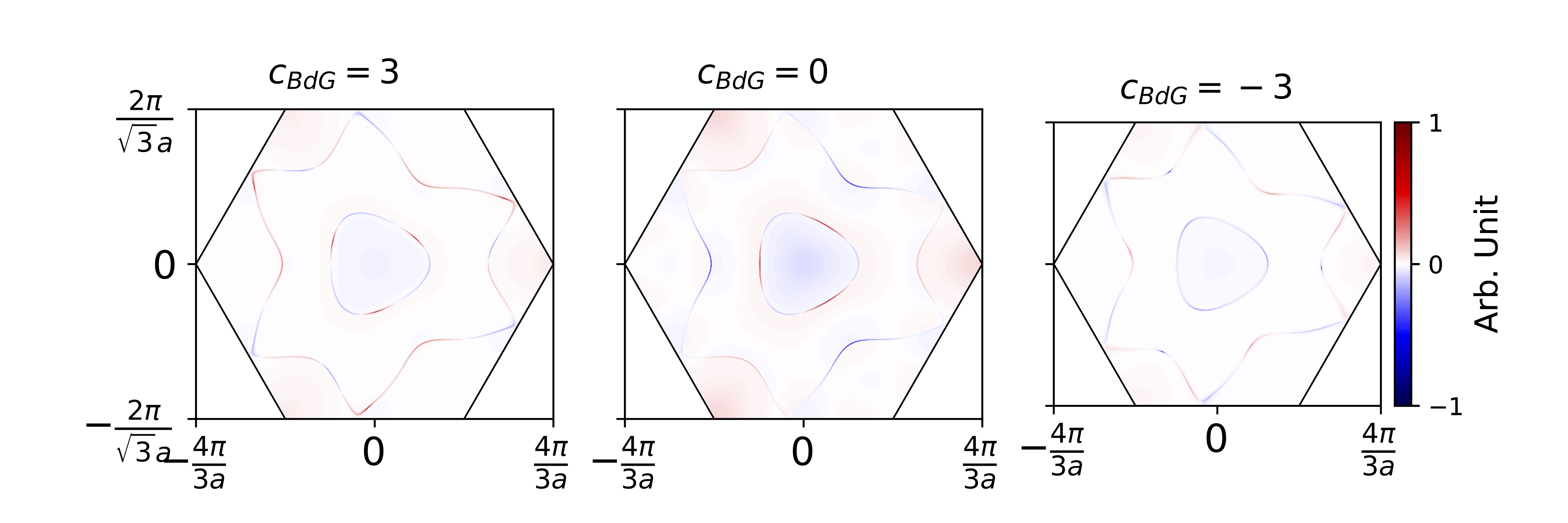}
\caption{Berry curvature for the three Ansatze with $c_{\text{BdG}}=+3,0,-3$.}\label{berry_C_spd}
\end{figure*}

The simplest ansatze with pairing along a vertical bond between $\mathrm{H}$ and $\mathrm{H}'$ always give $c_{\text{BdG}}=0$ (Two ansatze with only vertical pairings $\Delta_{\text{vert}}$ are given in Appendix.~\ref{app:vert_ans}). As we will be focusing on the relation between magnetization and $c_{\text{BdG}}$, we wish to obtain ansatze with nonzero Chern numbers of $\pm 3$. To find them, we consider ansatze with nearest-neighbor (NN) interlayer pairing as sketched in Fig.~\ref{fig:gap}(a). Furthermore, we construct a NN ansatz that closely approximates the solution of Eq.~\eqref{gapfunc}, see Fig.~\ref{fig:gap}(b). We thus obtain 
three ansatze with distinct Chern numbers $c_{\text{BdG}}=0,\pm3$:
\begin{itemize}
\item A $c_{\text{BdG}}=0$ ansatz:
\begin{equation}\label{paramswave}
\begin{aligned}
\Delta_1 &= \begin{pmatrix}
 \!-0.008 \!+\! 0.010i &0&0\\
0& \!-0.035\!-\!0.039i &  \!-0.033\!-\!0.009i\\
0& \!-0.033\!-\!0.009i &  \!-0.034\!+\!0.018i
\end{pmatrix},\\
\Delta_2
&=
\begin{pmatrix}\!-0.003\!-\!0.021i&0&0\\
0& \!-0.023\!+\!0.061i & 0.039\!-\!0.018i\\
0&0.039\!-\!0.018i& \!-0.049\!-\!0.017i
\end{pmatrix},
\end{aligned}
\end{equation}
\item A $c_{\text{BdG}}=3$ ansatz:
\begin{equation}\label{parampdwave}
\begin{aligned}
\Delta_1 &= \begin{pmatrix}
\!-0.932\!-\!1.258i& \!-0.852\!-\!0.992i&  0.02 \!-0.687i\\
  0.846\!-\!1.401i&0.248\!-\!0.20i & \!-0.552\!-\!2.172i\\
  0.396\!-\!0.707i&  0.202\!+\!1.598i&0.472\!+\!0.065i\end{pmatrix},\\
\Delta_2 &= \begin{pmatrix}0.389\!-\!0.334i& \!-1.076\!+\!0.292i& 0.107\!-\!1.187i\\ \!-0.002\!-\!0.706i &1.089\!+\!0.653i& 0.798\!+\!1.933j\\ \!-0.468\!+\!1.421i&  0.09 \!-\!0.151i &\!-1.435\!-\!0.928i\end{pmatrix}
\end{aligned}
\end{equation}
\item A $c_{\text{BdG}}=\!-\!3$ ansatz:
\begin{equation}\label{p3wave}
\begin{aligned}
\Delta_1 &= \begin{pmatrix}\!-0.566\!+\!1.234i &\!-0.021\!+\!0.600i&
       \!-0.921\!+\!0.495i\\  0.577\!+\!0.857i&
       \!-0.098\!+\!0.714i& \!-0.417\!-\!0.621i\\
       \!-0.484\!+\!1.283i& \!-2.129\!-\!0.199i&
       \!-0.0161\!-\!0.378i\end{pmatrix},\\
\Delta_2 &= \begin{pmatrix}\!-0.447\!-\!0.067i& 1.404\!-\!2.184i&
       \!-0.342\!-\!0.755i\\ \!-0.644\!+\!0.247i&
       \!-0.472\!+\!0.655i&  1.560\!+\!1.105i\\
        0.021\!+\!0.256i& \!-1.640\!+\!0.348i&
       \!-0.612\!+\!1.650i\end{pmatrix}
\end{aligned}      
\end{equation}
\end{itemize}
Their Berry curvatures in the BZ are plotted in Fig.~\ref{berry_C_spd}.

In the next subsection, we use the derived magnetization formula to calculate the magnetization of the above three interlayer pairing ansatze.

\subsection{Numerical results for magnetization}

\begin{figure}
\includegraphics[width=0.48\textwidth]{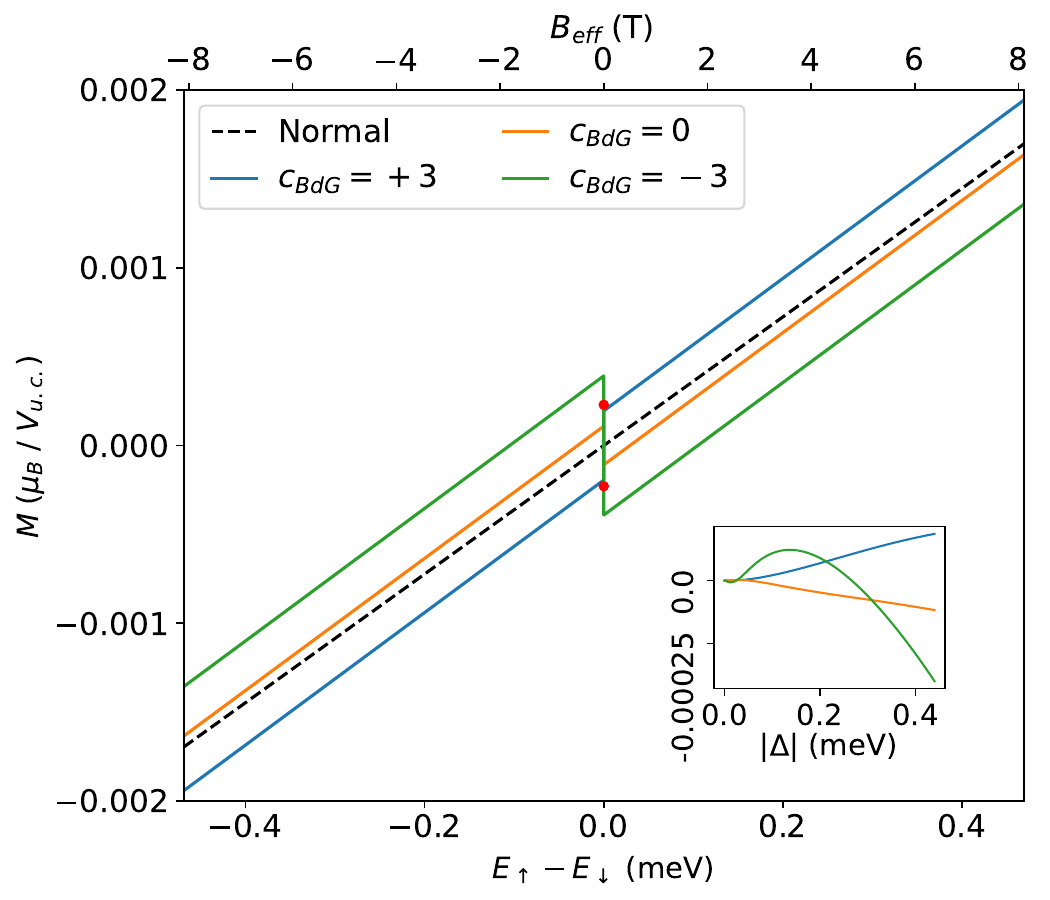}
\caption{Magnetization in the normal state (dashed line) and in the interlayer pairing ansatze (solid lines) with a gap of 0.44 meV. The three pairing ansatze have BdG Chern numbers $+3$, $0$, and $-3$. The lower and upper horizontal axes are related by $E_\uparrow-E_\downarrow = \mu_{\text{B}}B_{\text{eff}}$ (see the discussion above Eq.~\eqref{eq:mtot}). The two red dots denote the remnant magnetization inferred from the experiment \cite{persky2022magnetic}. Inset: magnetization as a function of the gap size for the three pairing ansatze. All calculations performed at $T=2$ K.}\label{mag}
\end{figure}

In Fig.~\ref{mag} we plot the magnetization as a function of the effective Zeeman energy splitting $E_\uparrow-E_\downarrow$ induced by the TRSB order parameter $\phi$ for the three NN ansatze mentioned above. All three exhibit a jump in the magnetization $\Delta M$ when $E_\uparrow-E_\downarrow$ crosses zero.
Only the $c_{\text{BdG}}=+3$ state gives a positive, paramagnetic jump in magnetization, while both the $c_{\text{BdG}}=0,-3$ ansatze give a negative (diamagnetic) jump. Comparison with the experimental hysteresis curve in Fig.~\ref{Phase_diagram} points at the $c_{\text{BdG}}\!=\!+3$ ansatz as the most promising candidate of the three ansatze. 
Encouragingly, the jump is of the same order of magnitude as the remnant field deduced experimentally, $M_{\text{SC,exp}}\!=\!2.28\times 10^{-4}\mu_{\text{B}}/V_{\text{u.c.}}$ (red dots in Fig.~\ref{mag}) \cite{persky2022magnetic}.

Further remarks on magnetization are in order. We notice from Fig.~\ref{mag} that the size of the hysteresis jump is approximately proportional to $c_{\text{BdG}}$, consistent with the understanding that a nonzero $c_{\text{BdG}}$ enhances the orbital magnetization, and further suggests that the orbital magnetization is the leading contribution to the total magnetization. However, we point out that the precise relation between the gap winding and the magnetization is not straightforward, as several factors (such as the gap sizes on the inner and outer FSs and temperature) can affect the magnetization, even for a given gap winding.
As an illustration to this caveat, we plot in the inset of Fig.~\ref{mag} the magnetization as a function of gap size at an infinitesimal field, and point out that the $c_{\text{BdG}}=-3$ ansatz can still have a positive magnetization at very small gap. Nevertheless, for the experimentally relevant temperature and gap size, the orbital magnetization is the leading contribution for large gap winding and the expected relation between the total gap winding and hysteresis jump holds (see Appendix \ref{app:SC_mag_further}). Combining the numerical results of $c_{\text{BdG}}$ and magnetization, 
we present the following picture for understanding the experimental observations on 4Hb-TaS$_2$:
if the $1\mathrm{H}$ and $1\mathrm{H'}$ layers are in an interlayer spin-triplet pairing state with $c_{\text{BdG}}=3$, and a large imbalance between the number of spin-up and spin-down Cooper pairs exists, an amplified magnetization can appear due to the excessive angular momentum carried by the majority spin Cooper pair.

\section{Conclusion and Discussion}\label{sec:con_dis}

In conclusion, we have formulated a phenomenological theory to understand the puzzling appearance of spontaneous vortices in 4Hb-TaS$_2$. Assuming that a weak FM order develops in the normal state, we showed that a weak coupling BCS instability exists in the interlayer, spin-polarized pairing channel. 
The breaking of TRS results in imbalance in the spin-up and spin-down pairings, and a total suppression of the minority spin pairing may be achieved, consistent with a single $T_c$ observed in experiment.
The angular momentum carried by the majority spin Cooper pair naturally enhances the magnetization and explains the appearance of spontaneous vortices in the superconducting phase. Our proposal of interlayer, spin-polarized pairing in a single spin component for 4Hb-TaS$_2$ is quite different from previous proposals \cite{MARGALIT2021168561, PhysRevB.103.224522,Konig}. Our proposed pairing state
can be verified in a spin-polarized STM experiment \cite{PhysRevLett.126.076802}, in which a gap should be observed  only in one spin component, but not in the other. We note that such a ``partial gap'' structure has been observed in a spin-unpolarized STM experiment \cite{nayak2021evidence}. More generally, our theory suggests a novel type of FM superconductor and could be relevant to a large family of centrosymmetric compounds \cite{fischer2022superconductivity}.

Finally, we mention a few predictions of our theory for existing and future experiments. 

First, we note that $T_c=2.7$ K is consistently reported in all experiments with or without a field training. 
It is plausible that the superconductivity in 4Hb-TaS$_2$ reported so far are all preceded by a TRSB state at higher temperature, and the existence of multiple randomly-oriented domains prohibits macroscopic ferromagnetism, hence the unobserved magnetic moment in the normal state. 

Second, in our proposed pairing state, the unpaired minority FS exhibits a linear-in-temperature specific heat, in agreement with existing experiment \cite{doi:10.1126/sciadv.aax9480}. We note that the specific heat reveals a $\sim 15\%$ residual contribution in the superconducting state \cite{doi:10.1126/sciadv.aax9480} and two distinct superconducting transitions are not observed down to $0.5$ K. Here we point out the possibility that superconductivity of the majority spin-species can promote enhanced fluctuations or pseudogap-like behavior for the minority spin-species.

Third, we have not specified the microscopic origin for the TRSB order parameter $\phi$.
While the most natural interpretation for $\phi$ is a FM order parameter, other interesting possibilities (e.g. $\phi$ being the scalar spin chirality in a CSL \cite{persky2022magnetic}) exist.

\begin{acknowledgments}
{\it Acknowledgements---}We thank J. E. Moore, Z. Dai, Y.-P. Lin, and H. Beidenkopf for helpful discussions. This research was supported by the Gordon and Betty Moore Foundation (C.L.) and a Simons Investigator award (E.A.).
S.C. was supported by the ARO through the MURI program (grant number W911NF17-1-0323). E.B. acknowledges support from the European Research Council (ERC) under grant HQMAT (grant agreement No. 817799), and the hospitality of the Aspen Center for Physics, supported by National Science Foundation grant PHY-2210452, where part of this work was done.
\end{acknowledgments}

\appendix

\section{Lattice Symmetry}

We set up the coordinates in such a way that $x$ axis is parallel to the bond $R_1$ in Fig.~\ref{fig:lattice}. Denote the lattice constant (i.e. the nearest neighbor bond distance) to be $a$. The origin is placed on an inversion center on the 1T layer. The $z$ periodicity is four layers (1H'-1T-1H-1T), with a four-layer distance $c$. We have $a=3.3381\text{ \AA}$ and $c = 23.728\text{ \AA}$ \cite{doi:10.1126/sciadv.aax9480}. The 4Hb-TaS$_2$ has a lattice symmetry of the No.~194 space group $P6_3/mmc$. The point group is $D_{6h}$ of order 24, generated by 
\begin{subequations}
\begin{align}
m_z&\colon\quad (x,y,z)\rightarrow (x,y,\pm \frac{1}{2}-z),\\
m_x&\colon\quad (x,y,z)\rightarrow (-x,y,z),\\
i&\colon\quad (x,y,z)\rightarrow (-x,-y,-z),\\
s_2&\colon\quad (x,y,z)\rightarrow (-x,-y,z+1/2),\\
c_3&\colon\quad (x,y,z) \rightarrow \Big(-\frac{x}{2}-\frac{\sqrt{3}}{2}y,\frac{\sqrt{3}}{2}x-\frac{1}{2}y,z\Big),
\end{align}
\end{subequations}

Symmetry action on the tight-binding Hamiltonian in momentum space gives \begin{subequations}
\begin{align}
m_z&\colon U^\dag_{m_z} \mathcal{H}_{\mathrm{H}}(\bm{k}) U_{m_z} = \mathcal{H}_{\mathrm{H}}(\bm{k}),\\
m_x& \colon U^\dag_{m_x} \mathcal{H}_{\mathrm{H}}(-k_x,k_y) U_{m_x} = \mathcal{H}_{\mathrm{H}}(\bm{k}),\\
i&\colon U^\dag_i \mathcal{H}_{\mathrm{H}}(-\bm{k}) U_i = \mathcal{H}_{\mathrm{H}}(\bm{k}),\\
c_3 & \colon U^\dag_{c_3} \mathcal{H}_{\mathrm{H}}(c_3^{-1}(\bm{k})) U_{c_3} = \mathcal{H}_{\mathrm{H}}(\bm{k}),\\
\mathcal{T} & \colon  \sigma^y \mathcal{H}_{\mathrm{H}}^*(-\bm{k}) \sigma^y = \mathcal{H}_{\mathrm{H}}(\bm{k}),
\end{align}
\end{subequations}
with
\begin{subequations}
\begin{align}
U_{m_z}&=    (-i \sigma^z)\otimes1_{3\times 3},\\
U_{m_x}&=   (-i \sigma^x)\otimes \mathrm{diag}(1,-1,1),\\
U_i &=  1_{2\times 2}\otimes 1_{3\times 3},\label{niv}\\
U_{c_3} &= e^{-i\frac{\pi}{3} \sigma^z} \otimes \mathbf{R} ,\\
U_{\mathcal{T}} & = i \sigma^y \otimes \mathcal{K} 1_{3\times 3},
\end{align}
\end{subequations}
where
\begin{equation}
\mathbf{R}\equiv  e^{-i\frac{2\pi}{3} L^z}= \begin{pmatrix} 1 &&\\& -1/2&-\sqrt{3}/2\\ & \sqrt{3}/2&-1/2\end{pmatrix}
\end{equation}
is the unitary matrix for the three-fold rotation $c_3$.

\section{Tight-binding Hamiltonian}\label{app:tb}

In angular momentum basis $|l,m\rangle$ we have $|d_{z^2}\rangle = |2,0\rangle$, $|d_{x^2-y^2}\rangle = \frac{1}{\sqrt{2}}(|2,2\rangle + |2,-2\rangle)$ and $|d_{xy}\rangle = -\frac{i}{\sqrt{2}} (|2,2\rangle -|2,-2\rangle)$. The angular momentum operators in the orbital subspace $\left(|d_{z^2}\rangle,|d_{xy}\rangle,|d_{x^2-y^2}\rangle\right)$ have the form
\begin{equation}\label{lzex}
L_z  = \left(\begin{array}{ccc}0&0&0\\0&0&2i \\ 0&-2i&0\end{array}\right),\quad {L}_x = {L}_y = 0.
\end{equation}

The nearest, second nearest and third nearest hopping matrices are constrained by lattice symmetry to the form
\begin{widetext}
\begin{equation}
R_1 = \left(\begin{array}{ccc}
t_0 & -t_1 & t_2 \\ t_1 & t_{11} & -t_{12} \\ t_2&t_{12}& t_{22} \end{array}\right),\quad
S_1 =  \left(\begin{array}{ccc}
r_0 & r_2 & - \frac{r_2}{\sqrt{3}} \\ r_1 & r_{11} & r_{12} \\
-\frac{r_1}{\sqrt{3}} & r_{12} & r_{11} + \frac{2}{\sqrt{3}}r_{12}\end{array}\right),\quad
T_1 =  \left(\begin{array}{ccc}
u_0 & -u_1 & u_2 \\ u_1 & u_{11} & - u_{12} \\ u_2 & u_{12} & u_{22}\end{array}\right).
\end{equation}
\end{widetext}
The tight-binding parameters are given in Table \ref{tbparams}.
We have 
\begin{equation}
\begin{aligned}
Q_2 = \mathbf{R} Q^\dag_1 \mathbf{R}^T,\quad
Q_3 = \mathbf{R}^T Q_1\mathbf{R} ,\quad
Q_4 = Q_1^\dag,\\
Q_5 = \mathbf{R} Q_1 \mathbf{R}^T,\quad
Q_6 = \mathbf{R}^T Q^\dag_1 \mathbf{R}\qquad\quad
\end{aligned}
\end{equation}
for $Q_i = R_i, S_i,T_i$ ($i=1,2,...,6$).

\begin{table}
\centering
\caption{Values of hopping and onsite parameters (units: meV) from Refs.~\cite{PhysRevB.98.144518,MARGALIT2021168561}.}\label{tbparams} 
\begin{tabular}{cccccc}
\hline
$t_0$& $t_1$ & $t_2$ & $t_{11}$ & $t_{12}$ & $t_{22}$ \\
$-0.1917$& $0.4057$ & $0.4367$ & $0.2739$ & $0.3608$ & $-0.1845$\\
\hline
$r_0$  & $r_1$ & $r_2$ & $r_{11}$ & $r_{12}$ & $r_{22}$\\
$0.0409$ & $-0.069$ & $0.0928$ & $-0.0066$ & $0.1116$ & $0$\\
\hline
$u_0$ & $u_1$ & $u_2$ & $u_{11}$ & $u_{12}$ & $u_{22}$ \\
$0.0405$ & $-0.0324$ & $-0.0141$ & $0.1205$ & $-0.0316$ & $-0.0778$\\
\hline
$\epsilon_0$ & $\epsilon_1$ & $\epsilon_2$ & $\mu_0$ & $\lambda_{\text{SO}}$ \\
$1.6507$ & $2.5703$ & $2.5703$ & $-0.0500$ & $0.1713$\\
\hline
\end{tabular}
\end{table}

Now we have $c_3\colon d_{\bm{k}} \rightarrow e^{-i\frac{2\pi}{3}L^z} e^{-i\frac{\pi}{3}\sigma} d_{c_3(\bm{k})} = e^{-i\frac{\pi}{3}\sigma}\mathbf{R}d_{c_3(\bm{k})}$, and we choose the gauge such that $u_{n\bm{k}} = \mathbf{R} u_{n,c_3(\bm{k})}$, therefore
\begin{equation}
c_3\colon c_{n\bm{k}}\rightarrow e^{-i\frac{\pi}{3}\sigma}c_{n,c_3(\bm{k})}.
\end{equation}

\section{Proposed phase diagram}

\begin{figure}[!thb]
\centering
\includegraphics[width=0.4\textwidth]{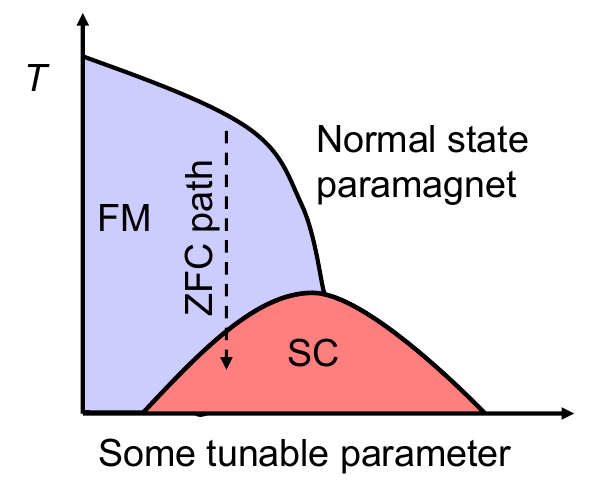}
\caption{Conjectured 2D phase diagram for 4Hb-TaS$_2$.}\label{phase:fig}
\end{figure}

In the main text we proposed a mechanism for the interlayer spin-triplet pairing mediated by the fluctuation of the TRSB/FM order parameter $\phi$. The underlying assumption for this mechanism is a two dimensional phase diagram spanned by temperature and another tunable parameter (such as doping, strain etc.), as shown in Fig.~\ref{phase:fig}. We assume that a quantum phase transition entering the FM phase happens at $T=0$ as the the horizontal axis parameter is tuned, around which the quantum fluctuation of the FM order is large; such a phase transition is hidden inside the dome of the SC phase. As a result, fluctuations of the order parameter $\phi$ may still be strong as one approaches along a low temperature horizontal path from the paramagnetic phase to the ferromagnetic phase. This path is distinct from the vertical, ZFC path traced out in the experiment, on which a FM order establishes at higher temperature and the fluctuation of $\phi$ in the FM phase is small.

\section{Further details for the gap equation and free energy}\label{app:gap_free_energy}

To numerically solve the BCS mean-field equation for the gap function, Eq.~3 in Main Text, we linearize it by substituting the Bogoliubov quasiparticle energy $\varepsilon_{\bm{k}}$ by the normal state quasiparticle energy $E_{\bm{k}}$; near the FSs we further have $E_{\bm{k}} = v_{\bm{k}}k_\perp$, where $k_\perp=|\bm{k}
_\perp|$ is the norm of the momentum $\bm{k}_\perp$ orthogonal to the Fermi surface tangent, $\bm{k}_\parallel$. We also write $d^2\bm{k} = dk_\perp dk_\parallel$. After integrating over $k_\perp$ and introducing a cutoff $\Lambda$ (of the order of the Fermi energy), we get
\begin{equation}\label{linearizedbcs}
\Delta_{\bm{k}}
= - V \log\left(\frac{\Lambda}{T_c}\right) \int_{\text{FSs}} \frac{d k'_\parallel}{(2\pi)^2 v_{\bm{k}'}} (\langle u_{\bm{k},\uparrow,\mathrm{H}}|u_{\bm{k}',\uparrow,\mathrm{H}}\rangle)^2 \Delta_{\bm{k}'},
\end{equation}
where $T_c$ is the transition temperature to be extracted from the solution of Eq.~\eqref{linearizedbcs}. Clearly, $T_c$ depends on the value of the effective attraction $V$; since we cannot estimate $V$ due to lack of enough experimental input, we will not attempt to extract the transition temperature. Our focus will be on the symmetry and topology of the gap function.

Eq.~\eqref{linearizedbcs} is then solved as an eigensystem equation. Define 
\begin{equation}
M_{\bm{k},\bm{k}'} = - \sqrt{ \frac{\Delta k \Delta k'}{v_{\bm{k}}v_{\bm{k}'}}} (\langle u_{\bm{k},\uparrow,\mathrm{H}}|u_{\bm{k}',\uparrow,\mathrm{H}}\rangle)^2,
\end{equation}
which is a symmetric matrix whose rows and columns are labeled by discretized momentum $\bm{k}$ that runs over the two FSs. The largest eigenvalue of $M$ gives the gap solution: denote the corresponding eigenvector as $a_{\bm{k}}$, then the gap function is obtained as 
\begin{equation}
\Delta_{\bm{k}} = \sqrt{\frac{v_{\bm{k}}}{\Delta k}} a_{\bm{k}}.
\end{equation}
The amplitude and phase of the solution $\Delta_{\bm{k}}$ is plotted in Fig.~3(b) in the main text.

\begin{figure*}
\centering
\includegraphics[width=0.8\textwidth]{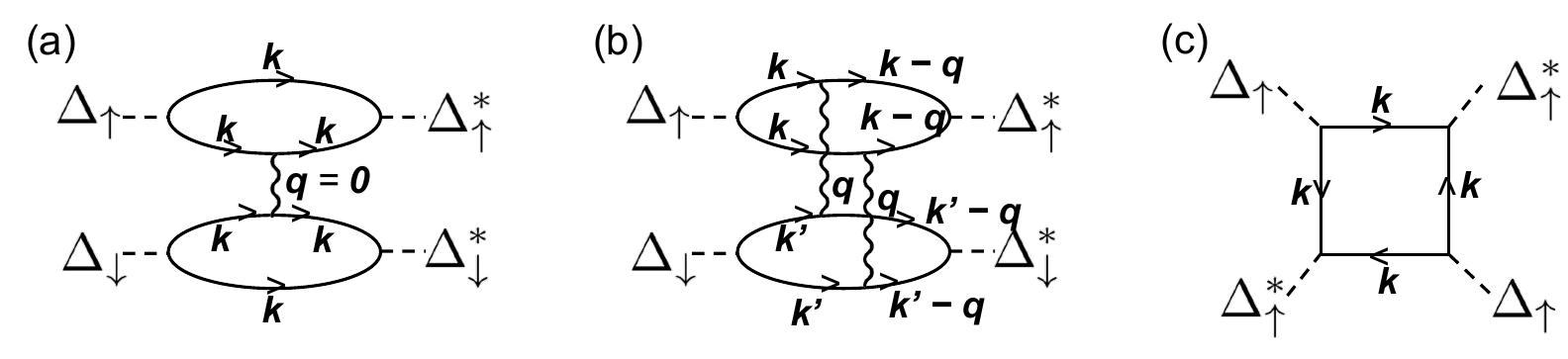}
\caption{Feynman diagrams at quartic level. The wavy and solid lines represent the propagators for the chiral field $\phi$ and electrons, respectively.}\label{fig:gl}
\end{figure*}

The quartic terms in the Ginzburg-Landau free energy is given in Eq.~\eqref{eq:f4}. Assuming a TRSB field $\phi$ that couples the electrons as an effective Zeeman field, $\phi(n_\uparrow-n_\downarrow)$, where $n_\sigma$ is the electron density for spin $\sigma$, the quartic term $v$ is produced by the diagrams in Fig.~\ref{fig:gl}(a)\&(b), and the term $u$ by the diagram in Fig.~\ref{fig:gl}(c). However, notice that in Fig.~\ref{fig:gl}(a) the TRSB field propagator carries zero momentum so this diagram gives a vanishing result. The diagram in Fig.~\ref{fig:gl}(b) does not have such a constraint on the TRSB field momentum $\bm{q}$ and it serves as the lowest order diagram contributing to the $v$ term. One concludes that $v\propto \chi^2$, where $\chi$ is the susceptibility of the TRSB field $\phi$.

\onecolumngrid

\section{Derivation of magnetization formula in an interlayer superconducting state}\label{app:mag_form_SC}

For the unitary matrix $U_{\text{BdG}}(\bm{k})$ that diagonalizes the BdG Hamiltonian in Eq.~\eqref{hbdgdiag}, we introduce the block notation
\begin{equation}
U_{\text{BdG}}(\bm{k})=
\left(\begin{array}{cc} [U11]& [U12] \\ {}[U21] & [U22]\end{array}\right).
\end{equation}
We have
\begin{equation}
H_{\text{BdG}} =\sum_{\bm{k}} \left(\sum_{i=1}^3\epsilon_i \gamma^\dag_{\bm{k}i}\gamma_{\bm{k}i}
+\sum_{i=4}^6 \epsilon_i \widetilde{\gamma}^\dag_{\bm{k}i}\widetilde{\gamma}_{\bm{k}i}\right),
\end{equation}
where 
\begin{equation}
\left(\begin{array}{c} \gamma_{\bm{k}} \\ \widetilde{\gamma}_{\bm{k}}\end{array}\right) = \left(\begin{array}{c} \gamma_{\bm{k}} \\ \left(\chi^\dag_{\bm{k}}\right)^T\end{array}\right) = U^\dag_{\text{BdG}}\left(\begin{array}{c} d_{\bm{k},\uparrow,\mathrm{H}}\\ \left(d^\dag_{-\bm{k},\uparrow,\mathrm{H}'}\right)^T\end{array}\right).
\end{equation}
We have 
\begin{subequations}
\begin{align}
u_{n\bm{k}}(\bm{r})
&=
\sum_{\bm{R},\ell} e^{i\bm{k}\cdot (\bm{R}-\bm{r})} u_{\bm{n}\ell}(\bm{k})\phi_{\ell \bm{R}}(\bm{r})=\sum_{\bm{R}} e^{i\bm{k}\cdot (\bm{R}-\bm{r})} w_{n\bm{R}}(\bm{r}),\\
w_{n\bm{R}}(\bm{r})
 &= \sum_{\bm{R}'\ell}\phi_{\ell \bm{R}'}(\bm{r})\frac{V_{\text{cell}}}{(2\pi)^3}\int_{\text{BZ}} e^{i\bm{k}\cdot (\bm{R}'-\bm{R})}u_{n\ell}(\bm{k}) d^3 k,\\
\bm{r}w_{n\bm{R}}(\bm{r}) &= \frac{V_{\text{cell}}}{(2\pi)^3}
\int_{\text{BZ}} [((-i)\partial_{\bm{k}}+\bm{R}) e^{i \bm{k}\cdot (\bm{r}-\bm{R})}]u_{n\bm{k}}(\bm{r}) d^3k,\label{rwnRr}
\end{align}
\end{subequations}
where $\phi_{\ell\bm{0}}$ is wavefunction for the atomic orbital $\ell$, and $u_{n\ell}(\bm{k})$ is the eigenvector; $u_{n\bm{k}}(\bm{r})$ is the Bloch wavefunction for the $n$th band, and $w_{n\bm{R}}(\bm{r})$ is the wavefunction for the Wannier orbital centered at site $\bm{R}$. Integrating Eq.~\eqref{rwnRr} by parts gives
\begin{equation}\label{rwRr}
\begin{aligned}
\bm{r}w_{n\bm{R}}(\bm{r}) &= \frac{V_{\text{cell}}}{(2\pi)^3}
\int_{\text{BZ}}  e^{i \bm{k}\cdot (\bm{r}-\bm{R})} (\bm{R}+i\partial_{\bm{k}})u_{n\bm{k}}(\bm{r}) d^3k\\
&=\frac{V_{\text{cell}}}{(2\pi)^3}
\int_{\text{BZ}} \sum_{\bm{R}',\ell} e^{i \bm{k}\cdot (\bm{R}'-\bm{R})} (\bm{R} - \bm{R}'+\bm{r}+i \partial_{\bm{k}})
u_{n\ell}(\bm{k})\phi_{\ell\bm{R}'}(\bm{r})d^3k.
\end{aligned}
\end{equation}
The second quantized angular momentum operator is defined as
\begin{equation}
\hat{L}_z = \int d \bm{r} \psi^\dag(\bm{r}) \hat{l}_z \psi(\bm{r}),
\end{equation}
where the quantum field annihilation operator is defined as 
\begin{equation}
\psi(\bm{r})
= \sum_{\bm{k}}\left(\sum_{i=1}^3 \langle \bm{r}|\gamma^\dag_{\bm{k}i}|0\rangle  \gamma_{\bm{k}i} + \langle \bm{r}|\gamma_{\bm{k}i}|0\rangle  \gamma^\dag_{\bm{k}i} + \sum_{i=4}^6 \langle \bm{r}|\widetilde{\gamma}^\dag_{\bm{k}i}|0\rangle  \widetilde{\gamma}_{\bm{k}i} + \langle \bm{r}|\widetilde{\gamma}_{\bm{k}i}|0\rangle  \widetilde{\gamma}^\dag_{\bm{k}i}\right).
\end{equation}
In the following, we first derive an expression for $\langle \hat{L}_z\rangle$ in terms of the Wannier orbitals, and then convert to tight-binding functions. Our derivation parallels that in Ref.~\cite{PhysRevB.101.134505}. We have
\begin{equation}
\begin{aligned}
\langle \hat{L}_z\rangle
&= 
 \sum_{\bm{k}}\int d \bm{r} \sum_{i=1}^3 \langle 0|\gamma_{\bm{k}i}|\bm{r}\rangle \hat{l}_z\langle \bm{r}|\gamma_{\bm{k}i}^\dag |0\rangle f_i
+
\langle 0|\gamma^\dag_{\bm{k}i}|\bm{r}\rangle \hat{l}_z\langle \bm{r}|\gamma_{\bm{k}i} |0\rangle (1-f_i)
+\sum_{i=4}^6 \langle 0|\widetilde{\gamma}_{\bm{k}i}|\bm{r}\rangle \hat{l}_z\langle \bm{r}|\widetilde{\gamma}_{\bm{k}i}^\dag |0\rangle f_i
+
\langle 0|\widetilde{\gamma}^\dag_{\bm{k}i}|\bm{r}\rangle \hat{l}_z\langle \bm{r}|\widetilde{\gamma}_{\bm{k}i} |0\rangle (1-f_i)\\
&=
\sum_{\bm{k}}\int d \bm{r} \sum_{i=1}^3 
\frac{1}{N} \sum_{\bm{R}',\bm{R}} e^{i \bm{k}\cdot (-\bm{R}'+\bm{R})}\sum_{j',j=1}^3\phi^*_{\bm{R}',\uparrow,\mathrm{H},j'}(\bm{r})[U11]^*_{j'i}
\hat{l}^{\mathrm{H}}_z\phi_{\bm{R},\uparrow,\mathrm{H},j}(\bm{r})[U11]_{ji} f_i\\
&\quad +
\sum_{\bm{k}}\int d \bm{r} \sum_{i=1}^3 
\frac{1}{N} \sum_{\bm{R}',\bm{R}} e^{i \bm{k}\cdot (\bm{R}'-\bm{R})}\sum_{j',j=1}^3\phi_{\bm{R}',\uparrow,\mathrm{H}',j'}(\bm{r})[U21]_{j'i}
\hat{l}^{\mathrm{H}'}_z\phi^*_{\bm{R},\uparrow,\mathrm{H'},j}(\bm{r})[U21]^*_{ji} (1-f_i)\\
&\quad +\sum_{\bm{k}}\int d \bm{r} \sum_{i=1}^3 
\frac{1}{N} \sum_{\bm{R}',\bm{R}} e^{i \bm{k}\cdot (-\bm{R}'+\bm{R})}\sum_{j',j=1}^3\phi^*_{\bm{R}',\uparrow,\mathrm{H},j'}(\bm{r})[U12]^*_{j'i}
\hat{l}^{\mathrm{H}}_z\phi_{\bm{R},\uparrow,\mathrm{H},j}(\bm{r})[U12]_{ji} f_{i+3}\\
&\quad +
\sum_{\bm{k}}\int d \bm{r} \sum_{i=1}^3 
\frac{1}{N} \sum_{\bm{R}',\bm{R}} e^{i \bm{k}\cdot (\bm{R}'-\bm{R})}\sum_{j',j=1}^3\phi_{\bm{R}',\uparrow,\mathrm{H}',j'}(\bm{r})[U22]_{j'i}
\hat{l}^{\mathrm{H}'}_z\phi^*_{\bm{R},\uparrow,\mathrm{H'},j}(\bm{r})[U22]^*_{ji}(1-f_{i+3})\\
&=
\sum_{\bm{k}}\int d \bm{r} \sum_{i=1}^3 
\frac{1}{N} \sum_{\bm{R}',\bm{R}} \left[e^{i \bm{k}\cdot (-\bm{R}'+\bm{R})}[W11]^*_{i,\bm{R}'}(\bm{r})
\hat{l}_z[W11]_{i,\bm{R}}(\bm{r}) f_i+
 e^{i \bm{k}\cdot (\bm{R}'-\bm{R})}[W21]_{i,\bm{R}'}
(\bm{r})\hat{l}_z [W21]_{i,\bm{R}}^*(\bm{r}) (1-f_i)\right]\\
&\quad +\sum_{\bm{k}}\int d \bm{r} \sum_{i=1}^3 
\frac{1}{N} \sum_{\bm{R}',\bm{R}}\left[ e^{i \bm{k}\cdot (-\bm{R}'+\bm{R})}[W12]^*_{i,\bm{R}'}(\bm{r})\hat{l}_z[W12]_{i,\bm{R}}(\bm{r}) f_{i+3}+
 e^{i \bm{k}\cdot (\bm{R}'-\bm{R})}[W22]_{i,\bm{R}'}(\bm{r})\hat{l}_z[W22]^*_{i,\bm{R}}(\bm{r})(1-f_{i+3})\right]\\
&= L_1 + L_2,
\end{aligned}
\end{equation}
where $n_i = \sum_{\bm{k}} f_{i,\bm{k}}$ is the density. We used $W$ to denote the Wannier orbitals. $L_1$ and $L_2$ comes from the decomposition of $\hat{l}_z$: by definition $\hat{l}_z = \bm{r}\times \bm{v} =  \bm{r}\times i \hat{H} (\bm{r}) \bm{r} = (\bm{r}-\bm{R}) \times i \hat{H}(\bm{r}) (\bm{r}-\bm{R}) + \bm{R}\times i \hat{H}(\bm{r})  (\bm{r}-\bm{R})$, the two terms respectively define $L_1$ and $L_2$. First, look at the first term: using Eq.~\eqref{rwRr} we have $(\bm{r}-\bm{R})[W](\bm{r})
= V_{\text{cell}}\int_{\text{BZ}} \frac{d^3k}{(2\pi)^3}
\sum_{\bm{R}',\ell}e^{i \bm{k}\cdot (\bm{R}'-\bm{R})}(-\bm{R}'+\bm{r}+i\partial_{\bm{k}})
u_{n\ell}(\bm{k}) \phi_{\ell \bm{R}'}(\bm{r})$, and
\begin{equation}
\begin{aligned}
L_1 &= 
i\sum_{\bm{k}}\int d \bm{r} \sum_{i,j',j} 
\frac{1}{N} \sum_{\bm{R}',\bm{R},\bm{R}'_1,\bm{R}_1}
V^2_{\text{cell}}\int_{\text{BZ}} \frac{d^3k_1}{(2\pi)^3}
e^{-i\bm{k}\cdot \bm{R}'} e^{-i \bm{k}'_1\cdot (\bm{R}'_1-\bm{R}')}(-\bm{R}'_1+\bm{r}-i\partial_{\bm{k}'_1})
[U11]^*_{j'i}(\bm{k}'_1) \phi^*_{\bm{R}'_1,\uparrow,\mathrm{H},j' }(\bm{r})\\
&\qquad \times H_{\mathrm{H}}(\bm{r})
\int_{\text{BZ}} \frac{d^3k_1}{(2\pi)^3}
e^{i\bm{k}\cdot \bm{R}} e^{i \bm{k}_1\cdot (\bm{R}_1-\bm{R})}(-\bm{R}_1+\bm{r}+i\partial_{\bm{k}_1})
[U11]_{ji}(\bm{k}_1) \phi_{\bm{R}_1,\uparrow,\mathrm{H},j}(\bm{r}) f_{i\bm{k}}\\
&\quad +i\sum_{\bm{k}}\int d \bm{r} \sum_{i,j',j'} 
\frac{1}{N} \sum_{\bm{R}',\bm{R},\bm{R}'_1,\bm{R}_1}
V^2_{\text{cell}}\int_{\text{BZ}} \frac{d^3k'_1}{(2\pi)^3}
e^{i\bm{k}\cdot \bm{R}'} e^{i \bm{k}'_1\cdot (\bm{R}'_1-\bm{R}')}(-\bm{R}'_1+\bm{r}+i\partial_{\bm{k}'_1})
[U21]_{j'i}(\bm{k}'_1) \phi^*_{\bm{R}'_1,\uparrow,\mathrm{H}',j'}(\bm{r})\\
&\qquad \times H_{\mathrm{H}'}(\bm{r})
\int_{\text{BZ}} \frac{d^3k_1}{(2\pi)^3}
e^{-i\bm{k}\cdot \bm{R}} e^{-i \bm{k}_1\cdot (\bm{R}_1-\bm{R})}(-\bm{R}'_1+\bm{r}-i\partial_{\bm{k}_1})
[U21]^*_{ji}(\bm{k}_1) \phi_{\bm{R}_1,\uparrow,\mathrm{H}',j'}(\bm{r}) (1-f_{i\bm{k}})\\
&\quad + \cdots,
\end{aligned}
\end{equation}
now the sums over $\bm{R}$ and $\bm{R}'$ can be done, which makes $\bm{k}=\bm{k}_1=\bm{k}'_1$. About overall factor: note that $\frac{1}{N}\sum_{\bm{k}} \rightarrow V_{\text{cell}} \int \frac{d^3k}{(2\pi)^3}$ which can be easily verified. So writing all the integral over $\bm{k}$ as sum, we have $\frac{1}{N^3}$. The sum over $\bm{R}$ gives one $N$ factor, and so does the sum over $\bm{R}'$, so we are left with $\frac{1}{N}\sum_{\bm{k}}\rightarrow V_{\text{cell}}\int \frac{d^3k}{(2\pi)^3}$. For convenience we denote
\begin{equation}
\int_{\bm{r},\bm{R},\bm{R}',\bm{k},i,j',j}\equiv \int d \bm{r} \sum_{i,j',j} 
\sum_{\bm{R}',\bm{R}}
V_{\text{cell}}\int_{\text{BZ}} \frac{d^3k}{(2\pi)^3},
\end{equation}
We rewrite $L_1$:
\begin{equation}
\begin{aligned}
L_1 &= 
i\int_{\bm{r},\bm{R},\bm{R}',\bm{k},i,j',j}
e^{i \bm{k}\cdot (\bm{R}-\bm{R}')}(-\bm{R}'+\bm{r})
[U11]^*_{j'i}(\bm{k}) \phi^*_{\bm{R}',\uparrow,\mathrm{H},j'}(\bm{r})\times H_{\mathrm{H}}(\bm{r})(-\bm{R}+\bm{r})
[U11]_{ji}(\bm{k}) \phi_{\bm{R},\uparrow,\mathrm{H},j}(\bm{r})f_{i\bm{k}}\\
&\quad +i\int_{\bm{r},\bm{R},\bm{R}',\bm{k},i,j',j}
e^{i \bm{k}\cdot (\bm{R}-\bm{R}')} 
\partial_{\bm{k}}
[U11]^*_{j'i}(\bm{k}) \phi^*_{\bm{R}',\uparrow,\mathrm{H},j'}(\bm{r})\times H_{\mathrm{H}}(\bm{r})\partial_{\bm{k}}
[U11]_{ji}(\bm{k}) \phi_{\bm{R},\uparrow,\mathrm{H},j}(\bm{r})f_{i\bm{k}}\\
&\quad +i\int_{\bm{r},\bm{R},\bm{R}',\bm{k},i,j',j}
e^{i \bm{k}\cdot (\bm{R}'-\bm{R})}(-\bm{R}'+\bm{r})
[U21]_{j'i}(\bm{k}) \phi_{\bm{R}',\uparrow,\mathrm{H}',j'}(\bm{r})\times H_{\mathrm{H}'}(\bm{r})(-\bm{R}+\bm{r})
[U21]^*_{ji}(\bm{k}) \phi^*_{\bm{R},\uparrow,\mathrm{H}',j}(\bm{r})(1-f_{i\bm{k}})\\
&\quad +i\int_{\bm{r},\bm{R},\bm{R}',\bm{k},i,j',j}
e^{i \bm{k}\cdot (\bm{R}'-\bm{R})} 
\partial_{\bm{k}}
[U21]_{j'i}(\bm{k}) \phi_{\bm{R}',\uparrow,\mathrm{H}',j'}(\bm{r})\times H_{\mathrm{H}'}(\bm{r})\partial_{\bm{k}}
[U21]^*_{ji}(\bm{k}) \phi^*_{\bm{R},\uparrow,\mathrm{H}',j}(\bm{r})(1-f_{i\bm{k}})\\
&\quad +\cdots,
\end{aligned}
\end{equation}
the first (and the third...) line gives the atomic angular momentum (and this sets $\bm{R}-\bm{R}'$) while the second (and the fourth...) line gives the Bloch angular momentum. Carrying out the integral over $\bm{r}$ gives:
$$\sum_{\bm{R}',\bm{R}} e^{i \bm{k}\cdot (\bm{R}-\bm{R}')} \int d\bm{r} \phi^*_{\bm{R}',\uparrow,\mathrm{H},j'}(\bm{r})\times H_{\mathrm{H}}(\bm{r}) \phi_{\bm{R},\uparrow,\mathrm{H},j}(\bm{r}) = 
\sum_{\bm{R}',\bm{R}}\int e^{i\bm{k}\cdot (\bm{R}-\bm{R}')}[H_{\mathrm{H},\bm{R}',\bm{R}}]_{j',j} = H_{\mathrm{H}}(\bm{k})_{j',j},
$$
$$\sum_{\bm{R}',\bm{R}} e^{i \bm{k}\cdot (\bm{R}'-\bm{R})} \int d\bm{r} \phi_{\bm{R}',\uparrow,\mathrm{H},j'}(\bm{r})\times H_{\mathrm{H}'}(\bm{r}) \phi^*_{\bm{R},\uparrow,\mathrm{H},j}(\bm{r}) = 
\sum_{\bm{R}',\bm{R}}\int e^{i\bm{k}\cdot (\bm{R}'-\bm{R})}[H_{\mathrm{H'},\bm{R}',\bm{R}}]_{j',j} = H_{\mathrm{H'}}(-\bm{k})_{j',j} ,
$$
where we have used the fact that all the $\phi_j$ are real (they are $d_{xy}$, $d_{x^2-y^2}$ and $d_{z^2}$ orbitals). Therefore, we have
\begin{equation}
\begin{aligned}
L_1&=\sum_{i,j',j} 
V_{\text{cell}}\int_{\text{BZ}} \frac{d^3k}{(2\pi)^3}
\left\{[U11]^*_{j'i}(\bm{k}) l^{\mathrm{H}}_{j'j}[U11]_{ji}(\bm{k}) +
i\partial_{\bm{k}}
[U11]^*_{j'i}(\bm{k}) \times H_{\mathrm{H}}(\bm{k})_{j'j}\partial_{\bm{k}}
[U11]_{'i}(\bm{k})\right\}f_{i\bm{k}}\\
&\quad +\sum_{i,j',j} 
V_{\text{cell}}\int_{\text{BZ}} \frac{d^3k}{(2\pi)^3}
\left\{[U21]_{j'i}(\bm{k}) l^{\mathrm{H}'}_{j'j}[U21]^*_{ji}(\bm{k}) + i
\partial_{\bm{k}}
[U21]_{j'i}(\bm{k}) \times H_{\mathrm{H}'}(-\bm{k})_{j'j}\partial_{\bm{k}}
[U21]^*_{ji}(\bm{k})\right\}(1-f_{i\bm{k}})\\ 
&\quad + ...\\
&=\sum_{i,j',j} 
V_{\text{cell}}\int_{\text{BZ}} \frac{d^3k}{(2\pi)^3}
\left\{[U11]^*_{j'i}(\bm{k}) l_{j'j}[U11]_{ji}(\bm{k}) +
i\partial_{\bm{k}}
[U11]^*_{j'i}(\bm{k}) \times H_{\mathrm{H}}(\bm{k})_{j'j}\partial_{\bm{k}}
[U11]_{'i}(\bm{k})\right\}f_{i\bm{k}}\\
&\quad +\sum_{i,j',j} 
V_{\text{cell}}\int_{\text{BZ}} \frac{d^3k}{(2\pi)^3}
\left\{-[U21]^*_{j'i}(\bm{k}) l_{j'j}[U21]_{ji}(\bm{k}) -i
\partial_{\bm{k}}
[U21]^*_{j'i}(\bm{k}) \times H^T_{\mathrm{H}}(\bm{k})_{j'j}\partial_{\bm{k}}
[U21]_{ji}(\bm{k})\right\}(1-f_{i\bm{k}})\\ 
&\quad +\sum_{i,j',j} 
V_{\text{cell}}\int_{\text{BZ}} \frac{d^3k}{(2\pi)^3}
\left\{[U12]^*_{j'i}(\bm{k}) l_{j'j}[U12]_{ji}(\bm{k}) +i
\partial_{\bm{k}}
[U12]^*_{j'i}(\bm{k}) \times H_{\mathrm{H}}(\bm{k})_{j'j}\partial_{\bm{k}}
[U12]_{ji}(\bm{k})\right\}f_{i+3,\bm{k}}\\ 
&\quad +\sum_{i,j',j} 
V_{\text{cell}}\int_{\text{BZ}} \frac{d^3k}{(2\pi)^3}
\left\{-[U22]^*_{j'i}(\bm{k}) l_{j'j}[U22]_{ji}(\bm{k}) -
i\partial_{\bm{k}}
[U22]^*_{j'i}(\bm{k}) \times H^T_{\mathrm{H}}(\bm{k})_{j'j}\partial_{\bm{k}}
[U22]_{ji}(\bm{k})\right\}(1-f_{i+3,\bm{k}}).
\end{aligned}
\end{equation}
We write the above as
\begin{equation}
L_1 = V_{\text{cell}}\int \frac{d^3 k}{(2\pi)^3}
U^\dag_{\text{BdG}}\left(1_{2\times 2}\otimes l\right) U_{\text{BdG}}
+
i \partial_{\bm{k}} U^\dag_{\text{BdG}} \left(\begin{array}{cc} \mathcal{H}_{\mathrm{H}}(\bm{k}) & \\ & - \mathcal{H}^T_{\mathrm{H}}(\bm{k})\end{array}\right)
\left(\begin{array}{ll} \partial_{\bm{k}}[U11](\bm{k}) f_{1:3,\bm{k}} & 
\partial_{\bm{k}} [U12](\bm{k}) f_{4:6,\bm{k}} \\
\partial_{\bm{k}}[U21](\bm{k}) (1-f_{1:3,\bm{k}}) & \partial_{\bm{k}} [U22](\bm{k}) (1-f_{4:6},\bm{k}) \end{array}\right),
\end{equation}
where $l$ is the $3\times 3$ matrix in the orbital basis, and $U_{\text{BdG}}=U_{\text{BdG}}(\bm{k})$. Then, note that magnetization is proportional to charge times angular momentum, so we define $M_{\text{LC}} = \frac{-eL_1}{V_{\text{cell}}}$ (note $e=|e|$ is the absolute value of the charge) and
\begin{equation}
M_{\text{LC}} = -e\;\mathrm{ReTr}
\int_{\text{BZ}} \frac{d^3k}{(2\pi)^3}
U^\dag_{\text{BdG}}\left(1_{2\times 2}\otimes l\right) U_{\text{BdG}}
+e\; \mathrm{ImTr}
\int_{\text{BZ}} \frac{d^3k}{(2\pi)^3}
\partial_{\bm{k}}
U^\dag_{\text{BdG}} \times \left(\mathcal{H}_{\text{BdG}}\Big|_{\Delta=0}\right)\left(\begin{array}{l} \partial_{\bm{k}}[U_{\text{BdG}}]_{1:3,:} f_{:,\bm{k}} \\\partial_{\bm{k}}[U_{\text{BdG}}]_{4:6,:} (1-f_{:,\bm{k}}) \end{array}\right).
\end{equation}
The \emph{overall} signs matches through in Eq. (10) and (11) of Ref.~\cite{PhysRevB.101.134505} (note that in \cite{PhysRevB.101.134505} $\gamma \propto -e$ carries a sign).

One can verify that when pairing term is zero the above formula correctly reduces to the normal state magnetization.

Then we have the second term $L_2$, that contains $\bm{R}\times i \hat{H}(\bm{r}) (\bm{r}-\bm{R})$. In parallel with \cite{PhysRevB.101.134505}, we propose that this term has the expression 
\begin{equation}
M_{\text{IC}}  = e\int \frac{d^3 k}{(2\pi)^3}
\mathrm{Im}\left\{\partial_{\bm{k}} U^\dag_{\text{BdG}} \times
\mathcal{E}_{:}\left(\begin{array}{l} \partial_{\bm{k}}[U_{\text{BdG}}]_{1:3,:} f_{:,\bm{k}} \\\partial_{\bm{k}}[U_{\text{BdG}}]_{4:6,:} (1-f_{:,\bm{k}}) \end{array}\right)\right\},
\end{equation}
where 
$\mathcal{E}_: = \mathcal{E} = \mathrm{diag}(\varepsilon _1,...,\varepsilon_6).$ This term is related to the Berry curvature of the BdG bands.

\section{Further details on magnetization}

\subsection{Normal state magnetization}

In 2D, the orbital magnetization $M_{\text{orb}}$ has the unit $\frac{e}{\hbar} \text{eV} = e \frac{\text{eV}}{6.582119569 \times 10^{-16} \text{eV}\cdot \text{s}} = 1.51927\times 10^{15} e\;\text{s}^{-1} =2.52939 \frac{\mu_B}{V_{\text{2D unit cell}}}$. 
Therefore, \begin{equation}\label{convertunit}
M_{\text{orb}} =\frac{\mu_B}{V_{\text{2D unit cell}}}\; 2.52939\times
\sum_n  \int \frac{d^2{k}}{(2\pi)^2}
\mathrm{Im}\langle \partial_{k_x} u_{n\bm{k}}| \mathsf{H}_{\bm{k}} + \mathsf{E}_{n \bm{k}}-2 \mathsf{E}_{\text{F}}|\partial_{k_y} u_{n\bm{k}}\rangle f_{n\bm{k}},
\end{equation}
where the sans-serif quantities denote the values when the unit is eV. Note that the orbital magnetization ~\eqref{normalorbmag} contains a localized angular momentum part: 
\begin{equation}
\begin{aligned}
M_{\text{orb,atom}}
&= \frac{-|e|}{2} g_{\text{orb}} \langle \hat{\bm{r}}\times \hat{\bm{v}}\rangle
= \frac{-|e|}{2m} g_{\text{orb}}\langle \hat{\bm{r}}\times m\hat{\bm{v}}\rangle
 = \frac{-|e|\hbar }{2m}g_{\text{orb}} \langle \hat{L}_z\rangle
= - g_{\text{orb}} \mu_B \langle \hat{L}_z\rangle\\
&= \frac{\mu_B}{V_{\text{2D unit cell}}} \cdot \left(-g_{\text{orb}} \frac{\sqrt{3}}{2}\right) \int \frac{d^2 \mathsf{k}}{(2\pi)^2} \sum_n u^\dag_{n\bm{k}} L_z u_{n\bm{k}}.
\end{aligned}
\end{equation}


The spin magnetization is
\begin{equation}
M_{\text{spin}}= \frac{1}{2} \mu_B g \langle c^\dag c\rangle
= \frac{\mu_B}{V_{\text{2D unit cell}}} \cdot \frac{\sqrt{3}}{2} \frac{1}{2} g \int \frac{d^2 \mathsf{k}}{(2\pi)^2} f_{n\bm{k}},
\end{equation}
where $\mathsf{k}$ is the numerical value that we use for momentum when the unit is $1/a$.

The normal state magnetization in the spin up sector is
\begin{equation}
(M^z_{\text{orb,t-b},\uparrow}, M^z_{\text{orb,atom},\uparrow},M^z_{\text{spin},\uparrow}) = (-0.0840,0.246,0.944)\; \mu_{\text{B}}/V_{\text{u.c.}}.
\end{equation}
We then calculate the total magnetization $M_\uparrow+M_\downarrow$ in presence of a magnetic field $B$. When $B=0$ we have $M_\uparrow+M_\downarrow=0$. When $B=5$ T, we have
\begin{equation}\label{eqorbtb}
(M^z_{\text{orb,t-b},\uparrow+\downarrow}, M^z_{\text{orb,atom},\uparrow+\downarrow},M^z_{\text{spin},\uparrow+\downarrow})
=(-0.000280, -0.000416, 0.00182)\;\mu_{\text{B}}/V_{\text{u.c.}}.
\end{equation}

\begin{table}\label{maganssd}
\caption{Magnetization for the three ansatze at temperatures $T=1$ K, 2 K, 5 K at gap size of $0.44$ meV and zero magnetic field. The magnetization is in the units of $10^{-4}\,\mu_B/V_{\text{u.c.}}$, where $V_{\text{u.c.}}$ is the volume of the four-layer unit cell of 4Hb-TaS$_2$.}
\centering
\begin{tabular}{c|ccc|ccc|ccc}
\hline
Ansatze&\multicolumn{3}{c|}{$c_{\text{BdG}}=+3$}&\multicolumn{3}{c|}{$c_{\text{BdG}}=0$}&\multicolumn{3}{c}{$c_{\text{BdG}}=-3$}\\
\hline
Magnetization&${M_{\text{orb,t-b},\uparrow+\downarrow}}$&$M_{\text{orb,atom},\uparrow+\downarrow}$&$M_{\text{spin},\uparrow+\downarrow}$ &${M_{\text{orb,t-b},\uparrow+\downarrow}}$&$M_{\text{orb,atom},\uparrow+\downarrow}$&$M_{\text{spin},\uparrow+\downarrow}$ &${M_{\text{orb,t-b},\uparrow+\downarrow}}$&$M_{\text{orb,atom},\uparrow+\downarrow}$&$M_{\text{spin},\uparrow+\downarrow}$ \\
\hline
$T=1$ K &$4.56$&$-1.52$&$3.04$&$-1.64$&$-5.48$&$6.92$&$-10.0$&$-0.604$&$2.18$\\
$T=2$ K &$3.58$&$-1.50$&$2.82$&$-2.58$&$-5.44$&$6.70$&$-10.9$&$-1.46$&$1.96$\\
$T=5$ K & $2.70$&$-1.56$&$2.22$&$-3.16$&$-5.52$&$5.94$&$-11.2$&$-0.238$&$10.1$\\
\hline
\end{tabular}
\end{table}

\subsection{Superconducting state magnetization: further plots}\label{app:SC_mag_further}

\begin{figure}[!thb]
\centering
\includegraphics[width=0.54\linewidth]{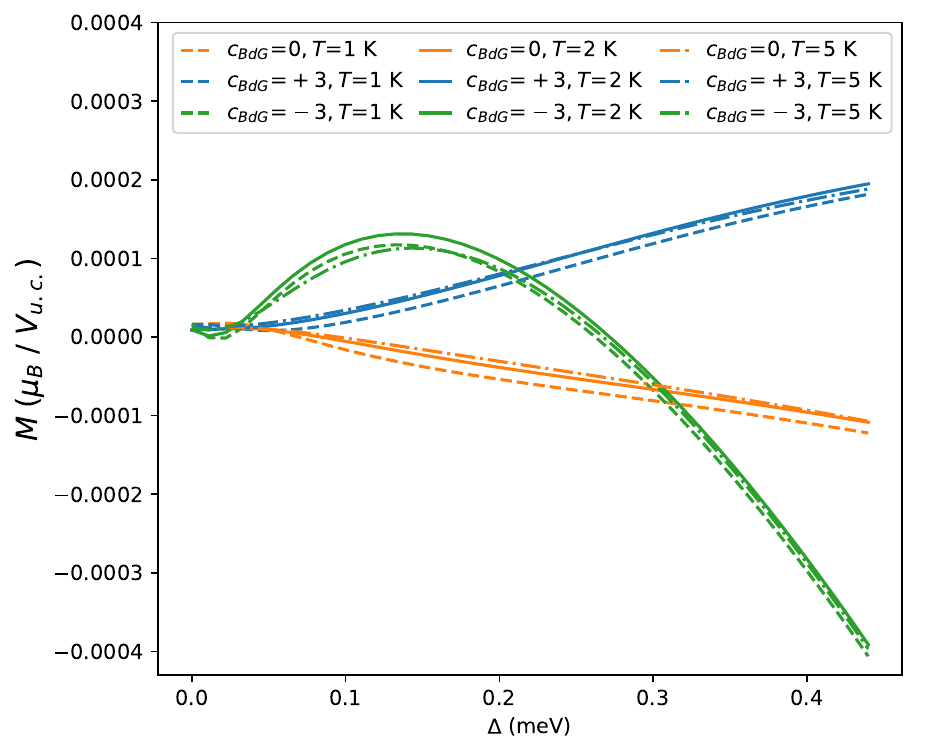}
\caption{Total magnetization $M$ as a function of the gap size $\Delta$ for the three ansatze with $c_{\text{BdG}} = +3,0,-3$ at zero field for three temperatures $T=1$ K, 2 K, 5 K.}\label{spd}
\end{figure}

We compute the total magnetization $M$ as a function of the gap size $\Delta$ using Eq.~\eqref{MM123}. The result is shown in Fig.~\eqref{spd}, for temperatures $T=1$ K ,2 K, and 5 K. 

For the realistic gap size $\Delta=0.44$ meV, we also compute the components of the magnetization, ${M_{\text{orb,t-b},\uparrow+\downarrow}}$, $M_{\text{orb,atom},\uparrow+\downarrow}$, $M_{\text{spin},\uparrow+\downarrow}$, at three temperatures $T=1$ K ,2 K, and 5 K. The result is given in Table \ref{maganssd}. 

\section{Pairing ansatze along a vertical bond}\label{app:vert_ans}

Here we consider a simple ``vertical'' pairing $\Delta_{\text{orb.}}$, i.e. the pairing exists only for a vertical bond between the $\mathrm{H}$ and $\mathrm{H'}$ layers. Note that $\Delta_{\text{orb.}}$ is a $3\times 3$ matrix in the orbital basis.

\begin{figure}[!thb]
\centering
\raisebox{3ex}{\includegraphics[width=0.15\textwidth]{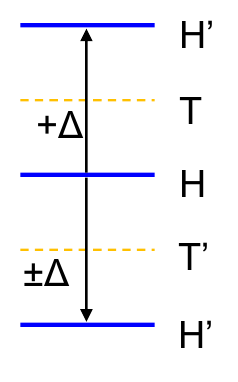}}\qquad\quad
\includegraphics[width=0.35\textwidth]{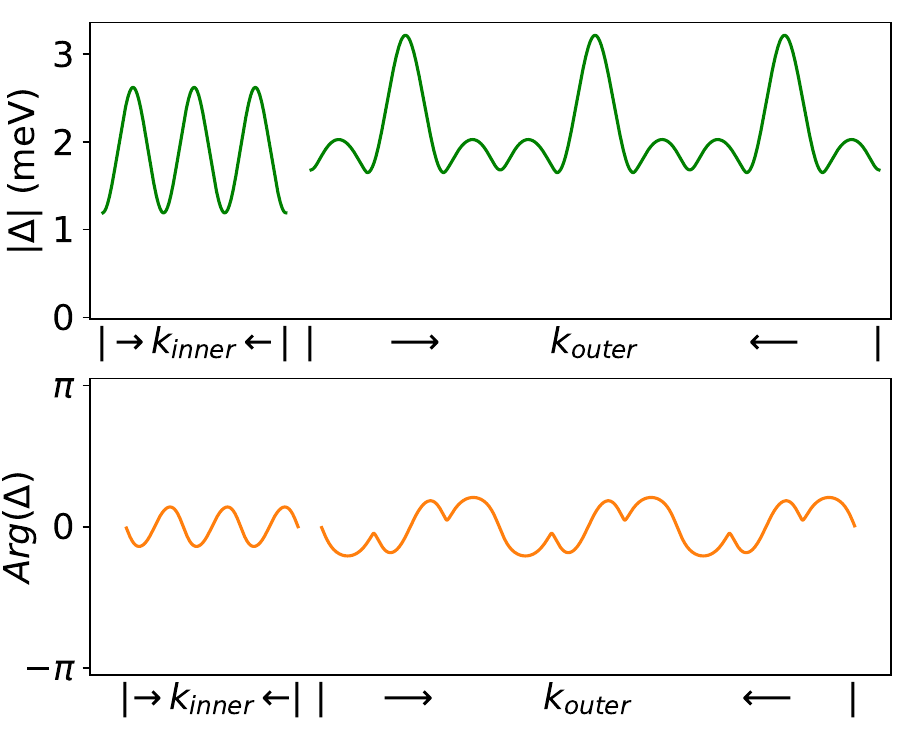}\qquad\quad
\includegraphics[width=0.35\textwidth]{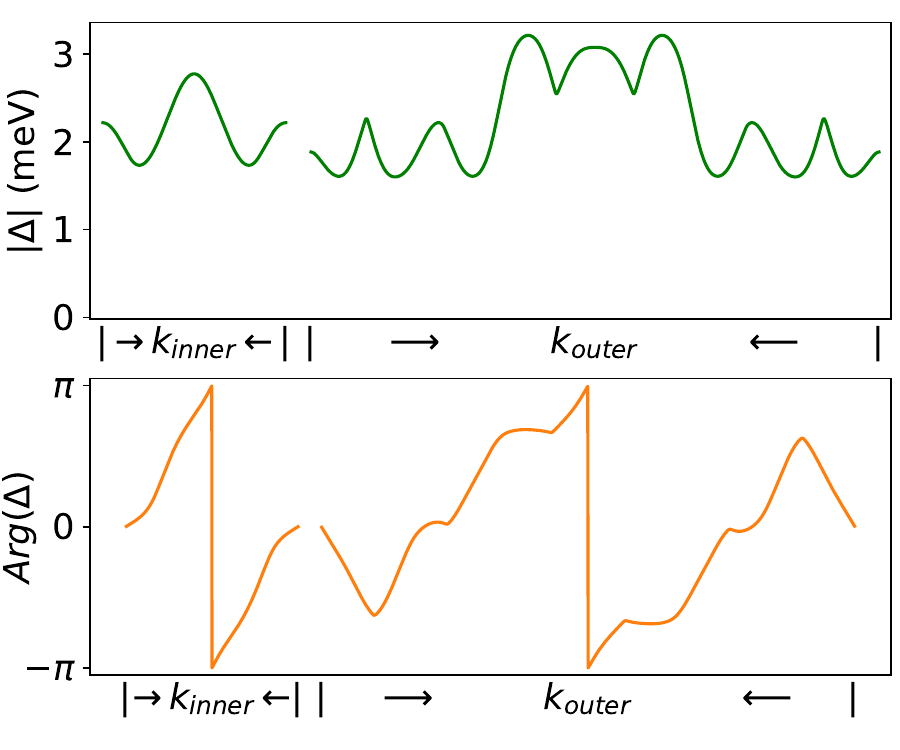}
\caption{Left: sketch of the vertical interlayer pairing.
Middle: Amplitude and winding of the gap function on the Fermi surfaces in the vertical pairing ansatz Eq.~\eqref{e1uans} (giving the $E_{\text{1u}}$ irrep), with zero gap winding on the FSs; Right: Amplitude and winding of the gap function on the Fermi surfaces in the vertical pairing ansatz Eq.~\eqref{e2uans} (giving the $E_{\text{2u}}$ irrep), with $2\pi$ gap winding on the FSs.}\label{e1uande2u}
\end{figure}

\begin{itemize}
\item The following gives a real space ansatz in the $E_{\text{u}}$ irrep, living on the $k_z=0$ layer:
\begin{equation}\label{e1uans}
\Delta_{\text{orb.},(\ell,\mathrm{H})\leftarrow(\ell,\mathrm{H'})}= 1_{3\times 3},\qquad
\Delta_{\text{orb.},(\ell,\mathrm{H})\leftarrow(\ell-1,\mathrm{H'})}  = -\Delta_{\text{orb.},(\ell,\mathrm{H})\leftarrow(\ell,\mathrm{H'})} 
\end{equation}
The winding of the gap function on the Fermi surface is zero, as verified in the middle panel of Fig.~\ref{e1uande2u}.
\item The following gives a real space ansatz in the $E_{\text{u}}$ irrep living on the $k_z=\pi$ layer:
\begin{equation}\label{e2uans}
\Delta_{\text{orb.},(\ell,\mathrm{H})\leftarrow(\ell,\mathrm{H'})}= \left(\begin{array}{ccc}0&0&1\\0&0&0\\1&0&0\end{array}\right),\qquad
\Delta_{\text{orb.},(\ell,\mathrm{H})\leftarrow(\ell-1,\mathrm{H'})}  =  \Delta_{\text{orb.},(\ell,\mathrm{H})\leftarrow(\ell,\mathrm{H'})} 
\end{equation}
The winding of the gap function on the Fermi surface is $2\pi$, as verified in the right panel of Fig.~\ref{e1uande2u}.
\end{itemize}
Note that both ansatze give a zero Chern number for the BdG band, due to the two fermi surface geometry (the inner FS is hole-like and the outer FS is electron like).

\begin{figure}[!thb]
\centering
\includegraphics[width=0.54\textwidth]{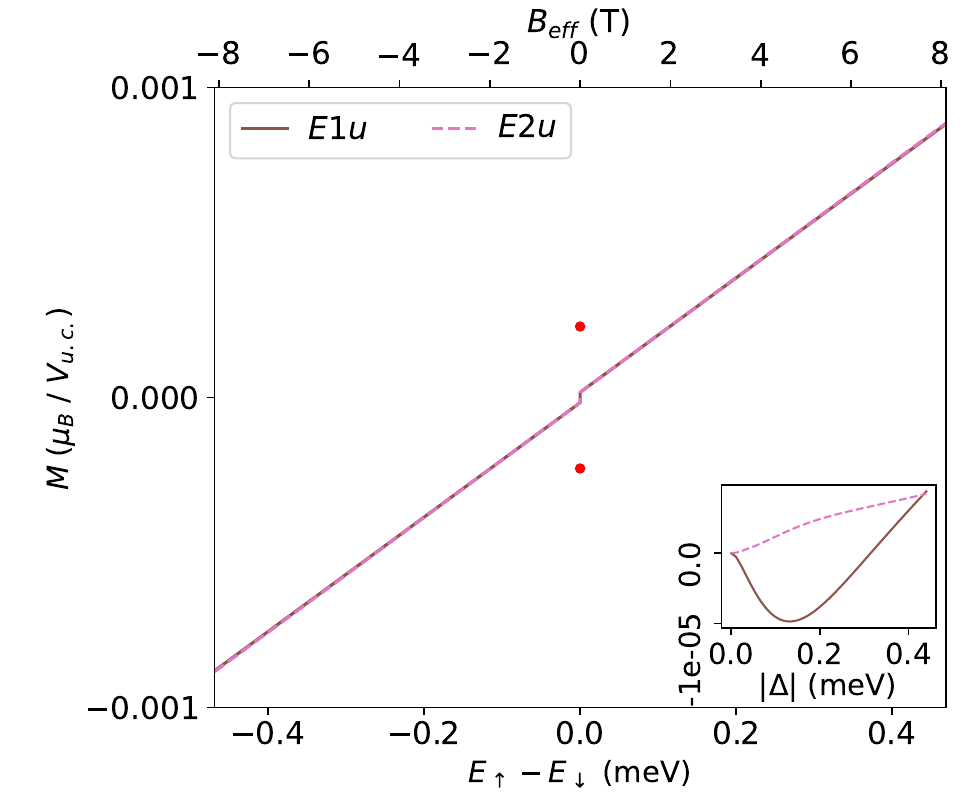}
\caption{Magnetization in the vertical pairing interlayer pairing ansatze \eqref{e1uans} and \eqref{e2uans} with a gap size of 0.44 meV. The lower and upper horizontal axes are related by $E_\uparrow-E_\downarrow = \mu_{\text{B}}B_{\text{eff}}$.The two red dots denote the remnant magnetization inferred from the experiment \cite{persky2022magnetic}. The inset shows the magnetization as a function of the gap size for the three pairing ansatze. Magnetizations are calculated at $T=2$ K.}\label{mag_vert}
\end{figure}

The magnetization as a function of $E_\uparrow-E_\downarrow$ induced by the TRSB order parameter $\phi$ for the two vertical pairing states Eq.~\eqref{e1uans} and Eq.~\eqref{e2uans} is shown in Fig.~\ref{mag_vert}.

\twocolumngrid

\bibliography{apssamp.bib}

\begin{thebibliography}{21}%
\makeatletter
\providecommand \@ifxundefined [1]{%
 \@ifx{#1\undefined}
}%
\providecommand \@ifnum [1]{%
 \ifnum #1\expandafter \@firstoftwo
 \else \expandafter \@secondoftwo
 \fi
}%
\providecommand \@ifx [1]{%
 \ifx #1\expandafter \@firstoftwo
 \else \expandafter \@secondoftwo
 \fi
}%
\providecommand \natexlab [1]{#1}%
\providecommand \enquote  [1]{``#1''}%
\providecommand \bibnamefont  [1]{#1}%
\providecommand \bibfnamefont [1]{#1}%
\providecommand \citenamefont [1]{#1}%
\providecommand \href@noop [0]{\@secondoftwo}%
\providecommand \href [0]{\begingroup \@sanitize@url \@href}%
\providecommand \@href[1]{\@@startlink{#1}\@@href}%
\providecommand \@@href[1]{\endgroup#1\@@endlink}%
\providecommand \@sanitize@url [0]{\catcode `\\12\catcode `\$12\catcode
  `\&12\catcode `\#12\catcode `\^12\catcode `\_12\catcode `\%12\relax}%
\providecommand \@@startlink[1]{}%
\providecommand \@@endlink[0]{}%
\providecommand \url  [0]{\begingroup\@sanitize@url \@url }%
\providecommand \@url [1]{\endgroup\@href {#1}{\urlprefix }}%
\providecommand \urlprefix  [0]{URL }%
\providecommand \Eprint [0]{\href }%
\providecommand \doibase [0]{https://doi.org/}%
\providecommand \selectlanguage [0]{\@gobble}%
\providecommand \bibinfo  [0]{\@secondoftwo}%
\providecommand \bibfield  [0]{\@secondoftwo}%
\providecommand \translation [1]{[#1]}%
\providecommand \BibitemOpen [0]{}%
\providecommand \bibitemStop [0]{}%
\providecommand \bibitemNoStop [0]{.\EOS\space}%
\providecommand \EOS [0]{\spacefactor3000\relax}%
\providecommand \BibitemShut  [1]{\csname bibitem#1\endcsname}%
\let\auto@bib@innerbib\@empty
\bibitem [{\citenamefont {Persky}\ \emph {et~al.}(2022)\citenamefont {Persky},
  \citenamefont {Bj{\o}rlig}, \citenamefont {Feldman}, \citenamefont
  {Almoalem}, \citenamefont {Altman}, \citenamefont {Berg}, \citenamefont
  {Kimchi}, \citenamefont {Ruhman}, \citenamefont {Kanigel},\ and\
  \citenamefont {Kalisky}}]{persky2022magnetic}%
  \BibitemOpen
  \bibfield  {author} {\bibinfo {author} {\bibfnamefont {E.}~\bibnamefont
  {Persky}}, \bibinfo {author} {\bibfnamefont {A.~V.}\ \bibnamefont
  {Bj{\o}rlig}}, \bibinfo {author} {\bibfnamefont {I.}~\bibnamefont {Feldman}},
  \bibinfo {author} {\bibfnamefont {A.}~\bibnamefont {Almoalem}}, \bibinfo
  {author} {\bibfnamefont {E.}~\bibnamefont {Altman}}, \bibinfo {author}
  {\bibfnamefont {E.}~\bibnamefont {Berg}}, \bibinfo {author} {\bibfnamefont
  {I.}~\bibnamefont {Kimchi}}, \bibinfo {author} {\bibfnamefont
  {J.}~\bibnamefont {Ruhman}}, \bibinfo {author} {\bibfnamefont
  {A.}~\bibnamefont {Kanigel}},\ and\ \bibinfo {author} {\bibfnamefont
  {B.}~\bibnamefont {Kalisky}},\ }\bibfield  {title} {\bibinfo {title}
  {Magnetic memory and spontaneous vortices in a van der waals
  superconductor},\ }\href@noop {} {\bibfield  {journal} {\bibinfo  {journal}
  {Nature}\ }\textbf {\bibinfo {volume} {607}},\ \bibinfo {pages} {692}
  (\bibinfo {year} {2022})}\BibitemShut {NoStop}%
\bibitem [{\citenamefont {Lin}(2022)}]{lin2022kondo}%
  \BibitemOpen
  \bibfield  {author} {\bibinfo {author} {\bibfnamefont {S.-Z.}\ \bibnamefont
  {Lin}},\ }\bibfield  {title} {\bibinfo {title} {Kondo enabled transmutation
  between spinons and superconducting vortices: origin of magnetic memory in
  {4Hb-TaS$_2$}},\ }\href@noop {} {\bibfield  {journal} {\bibinfo  {journal}
  {arXiv preprint arXiv:2210.06550}\ } (\bibinfo {year} {2022})}\BibitemShut
  {NoStop}%
\bibitem [{\citenamefont {Wilson}\ \emph {et~al.}(1975)\citenamefont {Wilson},
  \citenamefont {Di~Salvo},\ and\ \citenamefont {Mahajan}}]{wilson1975charge}%
  \BibitemOpen
  \bibfield  {author} {\bibinfo {author} {\bibfnamefont {J.~A.}\ \bibnamefont
  {Wilson}}, \bibinfo {author} {\bibfnamefont {F.}~\bibnamefont {Di~Salvo}},\
  and\ \bibinfo {author} {\bibfnamefont {S.}~\bibnamefont {Mahajan}},\
  }\bibfield  {title} {\bibinfo {title} {Charge-density waves and superlattices
  in the metallic layered transition metal dichalcogenides},\ }\href@noop {}
  {\bibfield  {journal} {\bibinfo  {journal} {Advances in Physics}\ }\textbf
  {\bibinfo {volume} {24}},\ \bibinfo {pages} {117} (\bibinfo {year}
  {1975})}\BibitemShut {NoStop}%
\bibitem [{\citenamefont {Kim}\ \emph {et~al.}(1994)\citenamefont {Kim},
  \citenamefont {Yamaguchi}, \citenamefont {Hasegawa},\ and\ \citenamefont
  {Kitazawa}}]{PhysRevLett.73.2103}%
  \BibitemOpen
  \bibfield  {author} {\bibinfo {author} {\bibfnamefont {J.-J.}\ \bibnamefont
  {Kim}}, \bibinfo {author} {\bibfnamefont {W.}~\bibnamefont {Yamaguchi}},
  \bibinfo {author} {\bibfnamefont {T.}~\bibnamefont {Hasegawa}},\ and\
  \bibinfo {author} {\bibfnamefont {K.}~\bibnamefont {Kitazawa}},\ }\bibfield
  {title} {\bibinfo {title} {Observation of mott localization gap using low
  temperature scanning tunneling spectroscopy in commensurate {1T-TaS$_2$}},\
  }\href {https://doi.org/10.1103/PhysRevLett.73.2103} {\bibfield  {journal}
  {\bibinfo  {journal} {Phys. Rev. Lett.}\ }\textbf {\bibinfo {volume} {73}},\
  \bibinfo {pages} {2103} (\bibinfo {year} {1994})}\BibitemShut {NoStop}%
\bibitem [{\citenamefont {Perfetti}\ \emph {et~al.}(2006)\citenamefont
  {Perfetti}, \citenamefont {Loukakos}, \citenamefont {Lisowski}, \citenamefont
  {Bovensiepen}, \citenamefont {Berger}, \citenamefont {Biermann},
  \citenamefont {Cornaglia}, \citenamefont {Georges},\ and\ \citenamefont
  {Wolf}}]{perfetti2006time}%
  \BibitemOpen
  \bibfield  {author} {\bibinfo {author} {\bibfnamefont {L.}~\bibnamefont
  {Perfetti}}, \bibinfo {author} {\bibfnamefont {P.}~\bibnamefont {Loukakos}},
  \bibinfo {author} {\bibfnamefont {M.}~\bibnamefont {Lisowski}}, \bibinfo
  {author} {\bibfnamefont {U.}~\bibnamefont {Bovensiepen}}, \bibinfo {author}
  {\bibfnamefont {H.}~\bibnamefont {Berger}}, \bibinfo {author} {\bibfnamefont
  {S.}~\bibnamefont {Biermann}}, \bibinfo {author} {\bibfnamefont
  {P.}~\bibnamefont {Cornaglia}}, \bibinfo {author} {\bibfnamefont
  {A.}~\bibnamefont {Georges}},\ and\ \bibinfo {author} {\bibfnamefont
  {M.}~\bibnamefont {Wolf}},\ }\bibfield  {title} {\bibinfo {title} {Time
  evolution of the electronic structure of {1T-TaS$_2$} through the
  insulator-metal transition},\ }\href@noop {} {\bibfield  {journal} {\bibinfo
  {journal} {Physical review letters}\ }\textbf {\bibinfo {volume} {97}},\
  \bibinfo {pages} {067402} (\bibinfo {year} {2006})}\BibitemShut {NoStop}%
\bibitem [{\citenamefont {Wang}\ \emph {et~al.}(2018)\citenamefont {Wang},
  \citenamefont {Sun}, \citenamefont {Abdelwahab}, \citenamefont {Cao},
  \citenamefont {Yu}, \citenamefont {Ju}, \citenamefont {Zhu}, \citenamefont
  {Fu}, \citenamefont {Chu}, \citenamefont {Xu} \emph
  {et~al.}}]{wang2018surface}%
  \BibitemOpen
  \bibfield  {author} {\bibinfo {author} {\bibfnamefont {Z.}~\bibnamefont
  {Wang}}, \bibinfo {author} {\bibfnamefont {Y.-Y.}\ \bibnamefont {Sun}},
  \bibinfo {author} {\bibfnamefont {I.}~\bibnamefont {Abdelwahab}}, \bibinfo
  {author} {\bibfnamefont {L.}~\bibnamefont {Cao}}, \bibinfo {author}
  {\bibfnamefont {W.}~\bibnamefont {Yu}}, \bibinfo {author} {\bibfnamefont
  {H.}~\bibnamefont {Ju}}, \bibinfo {author} {\bibfnamefont {J.}~\bibnamefont
  {Zhu}}, \bibinfo {author} {\bibfnamefont {W.}~\bibnamefont {Fu}}, \bibinfo
  {author} {\bibfnamefont {L.}~\bibnamefont {Chu}}, \bibinfo {author}
  {\bibfnamefont {H.}~\bibnamefont {Xu}}, \emph {et~al.},\ }\bibfield  {title}
  {\bibinfo {title} {Surface-limited superconducting phase transition on 1
  t-tas2},\ }\href@noop {} {\bibfield  {journal} {\bibinfo  {journal} {ACS
  nano}\ }\textbf {\bibinfo {volume} {12}},\ \bibinfo {pages} {12619} (\bibinfo
  {year} {2018})}\BibitemShut {NoStop}%
\bibitem [{\citenamefont {Gao}\ \emph {et~al.}(2020)\citenamefont {Gao},
  \citenamefont {Si}, \citenamefont {Luo}, \citenamefont {Yan}, \citenamefont
  {Jiang}, \citenamefont {Wang}, \citenamefont {Han}, \citenamefont {Tong},
  \citenamefont {Song}, \citenamefont {Zhu}, \citenamefont {Li}, \citenamefont
  {Lu},\ and\ \citenamefont {Sun}}]{PhysRevB.102.075138}%
  \BibitemOpen
  \bibfield  {author} {\bibinfo {author} {\bibfnamefont {J.~J.}\ \bibnamefont
  {Gao}}, \bibinfo {author} {\bibfnamefont {J.~G.}\ \bibnamefont {Si}},
  \bibinfo {author} {\bibfnamefont {X.}~\bibnamefont {Luo}}, \bibinfo {author}
  {\bibfnamefont {J.}~\bibnamefont {Yan}}, \bibinfo {author} {\bibfnamefont
  {Z.~Z.}\ \bibnamefont {Jiang}}, \bibinfo {author} {\bibfnamefont
  {W.}~\bibnamefont {Wang}}, \bibinfo {author} {\bibfnamefont {Y.~Y.}\
  \bibnamefont {Han}}, \bibinfo {author} {\bibfnamefont {P.}~\bibnamefont
  {Tong}}, \bibinfo {author} {\bibfnamefont {W.~H.}\ \bibnamefont {Song}},
  \bibinfo {author} {\bibfnamefont {X.~B.}\ \bibnamefont {Zhu}}, \bibinfo
  {author} {\bibfnamefont {Q.~J.}\ \bibnamefont {Li}}, \bibinfo {author}
  {\bibfnamefont {W.~J.}\ \bibnamefont {Lu}},\ and\ \bibinfo {author}
  {\bibfnamefont {Y.~P.}\ \bibnamefont {Sun}},\ }\bibfield  {title} {\bibinfo
  {title} {Origin of the large magnetoresistance in the candidate chiral
  superconductor $4{H}_{b}\text{\ensuremath{-}}\mathrm{Ta}{\mathrm{s}}_{2}$},\
  }\href {https://doi.org/10.1103/PhysRevB.102.075138} {\bibfield  {journal}
  {\bibinfo  {journal} {Phys. Rev. B}\ }\textbf {\bibinfo {volume} {102}},\
  \bibinfo {pages} {075138} (\bibinfo {year} {2020})}\BibitemShut {NoStop}%
\bibitem [{\citenamefont {Nayak}\ \emph {et~al.}(2021)\citenamefont {Nayak},
  \citenamefont {Steinbok}, \citenamefont {Roet}, \citenamefont {Koo},
  \citenamefont {Margalit}, \citenamefont {Feldman}, \citenamefont {Almoalem},
  \citenamefont {Kanigel}, \citenamefont {Fiete}, \citenamefont {Yan} \emph
  {et~al.}}]{nayak2021evidence}%
  \BibitemOpen
  \bibfield  {author} {\bibinfo {author} {\bibfnamefont {A.~K.}\ \bibnamefont
  {Nayak}}, \bibinfo {author} {\bibfnamefont {A.}~\bibnamefont {Steinbok}},
  \bibinfo {author} {\bibfnamefont {Y.}~\bibnamefont {Roet}}, \bibinfo {author}
  {\bibfnamefont {J.}~\bibnamefont {Koo}}, \bibinfo {author} {\bibfnamefont
  {G.}~\bibnamefont {Margalit}}, \bibinfo {author} {\bibfnamefont
  {I.}~\bibnamefont {Feldman}}, \bibinfo {author} {\bibfnamefont
  {A.}~\bibnamefont {Almoalem}}, \bibinfo {author} {\bibfnamefont
  {A.}~\bibnamefont {Kanigel}}, \bibinfo {author} {\bibfnamefont {G.~A.}\
  \bibnamefont {Fiete}}, \bibinfo {author} {\bibfnamefont {B.}~\bibnamefont
  {Yan}}, \emph {et~al.},\ }\bibfield  {title} {\bibinfo {title} {Evidence of
  topological boundary modes with topological nodal-point superconductivity},\
  }\href@noop {} {\bibfield  {journal} {\bibinfo  {journal} {Nature physics}\
  }\textbf {\bibinfo {volume} {17}},\ \bibinfo {pages} {1413} (\bibinfo {year}
  {2021})}\BibitemShut {NoStop}%
\bibitem [{\citenamefont {Nayak}\ \emph {et~al.}(2023)\citenamefont {Nayak},
  \citenamefont {Steinbok}, \citenamefont {Roet}, \citenamefont {Koo},
  \citenamefont {Feldman}, \citenamefont {Almoalem}, \citenamefont {Kanigel},
  \citenamefont {Yan}, \citenamefont {Rosch}, \citenamefont {Avraham},\ and\
  \citenamefont {Beidenkopf}}]{nayak2023first}%
  \BibitemOpen
  \bibfield  {author} {\bibinfo {author} {\bibfnamefont {A.~K.}\ \bibnamefont
  {Nayak}}, \bibinfo {author} {\bibfnamefont {A.}~\bibnamefont {Steinbok}},
  \bibinfo {author} {\bibfnamefont {Y.}~\bibnamefont {Roet}}, \bibinfo {author}
  {\bibfnamefont {J.}~\bibnamefont {Koo}}, \bibinfo {author} {\bibfnamefont
  {I.}~\bibnamefont {Feldman}}, \bibinfo {author} {\bibfnamefont
  {A.}~\bibnamefont {Almoalem}}, \bibinfo {author} {\bibfnamefont
  {A.}~\bibnamefont {Kanigel}}, \bibinfo {author} {\bibfnamefont
  {B.}~\bibnamefont {Yan}}, \bibinfo {author} {\bibfnamefont {A.}~\bibnamefont
  {Rosch}}, \bibinfo {author} {\bibfnamefont {N.}~\bibnamefont {Avraham}},\
  and\ \bibinfo {author} {\bibfnamefont {H.}~\bibnamefont {Beidenkopf}},\
  }\bibfield  {title} {\bibinfo {title} {First-order quantum phase transition
  in the hybrid metal--mott insulator transition metal dichalcogenide
  {4Hb-TaS$_2$}},\ }\href {https://doi.org/10.1073/pnas.2304274120} {\bibfield
  {journal} {\bibinfo  {journal} {Proceedings of the National Academy of
  Sciences}\ }\textbf {\bibinfo {volume} {120}},\ \bibinfo {pages}
  {e2304274120} (\bibinfo {year} {2023})},\ \Eprint
  {https://arxiv.org/abs/https://www.pnas.org/doi/pdf/10.1073/pnas.2304274120}
  {https://www.pnas.org/doi/pdf/10.1073/pnas.2304274120} \BibitemShut {NoStop}%
\bibitem [{\citenamefont {Ribak}\ \emph {et~al.}(2020)\citenamefont {Ribak},
  \citenamefont {Skiff}, \citenamefont {Mograbi}, \citenamefont {Rout},
  \citenamefont {Fischer}, \citenamefont {Ruhman}, \citenamefont {Chashka},
  \citenamefont {Dagan},\ and\ \citenamefont
  {Kanigel}}]{doi:10.1126/sciadv.aax9480}%
  \BibitemOpen
  \bibfield  {author} {\bibinfo {author} {\bibfnamefont {A.}~\bibnamefont
  {Ribak}}, \bibinfo {author} {\bibfnamefont {R.~M.}\ \bibnamefont {Skiff}},
  \bibinfo {author} {\bibfnamefont {M.}~\bibnamefont {Mograbi}}, \bibinfo
  {author} {\bibfnamefont {P.~K.}\ \bibnamefont {Rout}}, \bibinfo {author}
  {\bibfnamefont {M.~H.}\ \bibnamefont {Fischer}}, \bibinfo {author}
  {\bibfnamefont {J.}~\bibnamefont {Ruhman}}, \bibinfo {author} {\bibfnamefont
  {K.}~\bibnamefont {Chashka}}, \bibinfo {author} {\bibfnamefont
  {Y.}~\bibnamefont {Dagan}},\ and\ \bibinfo {author} {\bibfnamefont
  {A.}~\bibnamefont {Kanigel}},\ }\bibfield  {title} {\bibinfo {title} {Chiral
  superconductivity in the alternate stacking compound {4Hb-TaS$_2$}},\ }\href
  {https://doi.org/10.1126/sciadv.aax9480} {\bibfield  {journal} {\bibinfo
  {journal} {Science Advances}\ }\textbf {\bibinfo {volume} {6}},\ \bibinfo
  {pages} {eaax9480} (\bibinfo {year} {2020})},\ \Eprint
  {https://arxiv.org/abs/https://www.science.org/doi/pdf/10.1126/sciadv.aax9480}
  {https://www.science.org/doi/pdf/10.1126/sciadv.aax9480} \BibitemShut
  {NoStop}%
\bibitem [{\citenamefont {Liu}\ \emph {et~al.}(2013)\citenamefont {Liu},
  \citenamefont {Shan}, \citenamefont {Yao}, \citenamefont {Yao},\ and\
  \citenamefont {Xiao}}]{liu2013three}%
  \BibitemOpen
  \bibfield  {author} {\bibinfo {author} {\bibfnamefont {G.-B.}\ \bibnamefont
  {Liu}}, \bibinfo {author} {\bibfnamefont {W.-Y.}\ \bibnamefont {Shan}},
  \bibinfo {author} {\bibfnamefont {Y.}~\bibnamefont {Yao}}, \bibinfo {author}
  {\bibfnamefont {W.}~\bibnamefont {Yao}},\ and\ \bibinfo {author}
  {\bibfnamefont {D.}~\bibnamefont {Xiao}},\ }\bibfield  {title} {\bibinfo
  {title} {Three-band tight-binding model for monolayers of group-vib
  transition metal dichalcogenides},\ }\href@noop {} {\bibfield  {journal}
  {\bibinfo  {journal} {Physical Review B}\ }\textbf {\bibinfo {volume} {88}},\
  \bibinfo {pages} {085433} (\bibinfo {year} {2013})}\BibitemShut {NoStop}%
\bibitem [{\citenamefont {Margalit}\ \emph {et~al.}(2021)\citenamefont
  {Margalit}, \citenamefont {Berg},\ and\ \citenamefont
  {Oreg}}]{MARGALIT2021168561}%
  \BibitemOpen
  \bibfield  {author} {\bibinfo {author} {\bibfnamefont {G.}~\bibnamefont
  {Margalit}}, \bibinfo {author} {\bibfnamefont {E.}~\bibnamefont {Berg}},\
  and\ \bibinfo {author} {\bibfnamefont {Y.}~\bibnamefont {Oreg}},\ }\bibfield
  {title} {\bibinfo {title} {Theory of multi-orbital topological
  superconductivity in transition metal dichalcogenides},\ }\href
  {https://doi.org/https://doi.org/10.1016/j.aop.2021.168561} {\bibfield
  {journal} {\bibinfo  {journal} {Annals of Physics}\ }\textbf {\bibinfo
  {volume} {435}},\ \bibinfo {pages} {168561} (\bibinfo {year} {2021})},\
  \bibinfo {note} {special issue on Philip W. Anderson}\BibitemShut {NoStop}%
\bibitem [{\citenamefont {Ceresoli}\ \emph {et~al.}(2006)\citenamefont
  {Ceresoli}, \citenamefont {Thonhauser}, \citenamefont {Vanderbilt},\ and\
  \citenamefont {Resta}}]{PhysRevB.74.024408}%
  \BibitemOpen
  \bibfield  {author} {\bibinfo {author} {\bibfnamefont {D.}~\bibnamefont
  {Ceresoli}}, \bibinfo {author} {\bibfnamefont {T.}~\bibnamefont
  {Thonhauser}}, \bibinfo {author} {\bibfnamefont {D.}~\bibnamefont
  {Vanderbilt}},\ and\ \bibinfo {author} {\bibfnamefont {R.}~\bibnamefont
  {Resta}},\ }\bibfield  {title} {\bibinfo {title} {Orbital magnetization in
  crystalline solids: Multi-band insulators, chern insulators, and metals},\
  }\href {https://doi.org/10.1103/PhysRevB.74.024408} {\bibfield  {journal}
  {\bibinfo  {journal} {Phys. Rev. B}\ }\textbf {\bibinfo {volume} {74}},\
  \bibinfo {pages} {024408} (\bibinfo {year} {2006})}\BibitemShut {NoStop}%
\bibitem [{web()}]{website}%
  \BibitemOpen
  \bibinfo {note} {{SQUID sensitivity taken from
  \url{https://beenalab.biu.ac.il/}.}}\BibitemShut {Stop}%
\bibitem [{\citenamefont {Shi}\ \emph {et~al.}(2007)\citenamefont {Shi},
  \citenamefont {Vignale}, \citenamefont {Xiao},\ and\ \citenamefont
  {Niu}}]{PhysRevLett.99.197202}%
  \BibitemOpen
  \bibfield  {author} {\bibinfo {author} {\bibfnamefont {J.}~\bibnamefont
  {Shi}}, \bibinfo {author} {\bibfnamefont {G.}~\bibnamefont {Vignale}},
  \bibinfo {author} {\bibfnamefont {D.}~\bibnamefont {Xiao}},\ and\ \bibinfo
  {author} {\bibfnamefont {Q.}~\bibnamefont {Niu}},\ }\bibfield  {title}
  {\bibinfo {title} {Quantum theory of orbital magnetization and its
  generalization to interacting systems},\ }\href
  {https://doi.org/10.1103/PhysRevLett.99.197202} {\bibfield  {journal}
  {\bibinfo  {journal} {Phys. Rev. Lett.}\ }\textbf {\bibinfo {volume} {99}},\
  \bibinfo {pages} {197202} (\bibinfo {year} {2007})}\BibitemShut {NoStop}%
\bibitem [{\citenamefont {Dentelski}\ \emph {et~al.}(2021)\citenamefont
  {Dentelski}, \citenamefont {Day-Roberts}, \citenamefont {Birol},
  \citenamefont {Fernandes},\ and\ \citenamefont
  {Ruhman}}]{PhysRevB.103.224522}%
  \BibitemOpen
  \bibfield  {author} {\bibinfo {author} {\bibfnamefont {D.}~\bibnamefont
  {Dentelski}}, \bibinfo {author} {\bibfnamefont {E.}~\bibnamefont
  {Day-Roberts}}, \bibinfo {author} {\bibfnamefont {T.}~\bibnamefont {Birol}},
  \bibinfo {author} {\bibfnamefont {R.~M.}\ \bibnamefont {Fernandes}},\ and\
  \bibinfo {author} {\bibfnamefont {J.}~\bibnamefont {Ruhman}},\ }\bibfield
  {title} {\bibinfo {title} {Robust gapless superconductivity in
  {4Hb-TaS$_2$}},\ }\href {https://doi.org/10.1103/PhysRevB.103.224522}
  {\bibfield  {journal} {\bibinfo  {journal} {Phys. Rev. B}\ }\textbf {\bibinfo
  {volume} {103}},\ \bibinfo {pages} {224522} (\bibinfo {year}
  {2021})}\BibitemShut {NoStop}%
\bibitem [{\citenamefont {K\"onig}(2024)}]{Konig}%
  \BibitemOpen
  \bibfield  {author} {\bibinfo {author} {\bibfnamefont {E.~J.}\ \bibnamefont
  {K\"onig}},\ }\bibfield  {title} {\bibinfo {title} {Type-ii heavy fermi
  liquids and the magnetic memory of
  $4hb\text{\ensuremath{-}}{\mathrm{tas}}_{2}$},\ }\href
  {https://doi.org/10.1103/PhysRevResearch.6.L012058} {\bibfield  {journal}
  {\bibinfo  {journal} {Phys. Rev. Res.}\ }\textbf {\bibinfo {volume} {6}},\
  \bibinfo {pages} {L012058} (\bibinfo {year} {2024})}\BibitemShut {NoStop}%
\bibitem [{\citenamefont {Wang}\ \emph {et~al.}(2021)\citenamefont {Wang},
  \citenamefont {Wiebe}, \citenamefont {Zhong}, \citenamefont {Gu},\ and\
  \citenamefont {Wiesendanger}}]{PhysRevLett.126.076802}%
  \BibitemOpen
  \bibfield  {author} {\bibinfo {author} {\bibfnamefont {D.}~\bibnamefont
  {Wang}}, \bibinfo {author} {\bibfnamefont {J.}~\bibnamefont {Wiebe}},
  \bibinfo {author} {\bibfnamefont {R.}~\bibnamefont {Zhong}}, \bibinfo
  {author} {\bibfnamefont {G.}~\bibnamefont {Gu}},\ and\ \bibinfo {author}
  {\bibfnamefont {R.}~\bibnamefont {Wiesendanger}},\ }\bibfield  {title}
  {\bibinfo {title} {Spin-polarized yu-shiba-rusinov states in an iron-based
  superconductor},\ }\href {https://doi.org/10.1103/PhysRevLett.126.076802}
  {\bibfield  {journal} {\bibinfo  {journal} {Phys. Rev. Lett.}\ }\textbf
  {\bibinfo {volume} {126}},\ \bibinfo {pages} {076802} (\bibinfo {year}
  {2021})}\BibitemShut {NoStop}%
\bibitem [{\citenamefont {Fischer}\ \emph {et~al.}(2023)\citenamefont
  {Fischer}, \citenamefont {Sigrist}, \citenamefont {Agterberg},\ and\
  \citenamefont {Yanase}}]{fischer2022superconductivity}%
  \BibitemOpen
  \bibfield  {author} {\bibinfo {author} {\bibfnamefont {M.~H.}\ \bibnamefont
  {Fischer}}, \bibinfo {author} {\bibfnamefont {M.}~\bibnamefont {Sigrist}},
  \bibinfo {author} {\bibfnamefont {D.~F.}\ \bibnamefont {Agterberg}},\ and\
  \bibinfo {author} {\bibfnamefont {Y.}~\bibnamefont {Yanase}},\ }\bibfield
  {title} {\bibinfo {title} {Superconductivity and local inversion-symmetry
  breaking},\ }\href
  {https://doi.org/https://doi.org/10.1146/annurev-conmatphys-040521-042511}
  {\bibfield  {journal} {\bibinfo  {journal} {Annual Review of Condensed Matter
  Physics}\ }\textbf {\bibinfo {volume} {14}},\ \bibinfo {pages} {153}
  (\bibinfo {year} {2023})}\BibitemShut {NoStop}%
\bibitem [{\citenamefont {M\"ockli}\ and\ \citenamefont
  {Khodas}(2018)}]{PhysRevB.98.144518}%
  \BibitemOpen
  \bibfield  {author} {\bibinfo {author} {\bibfnamefont {D.}~\bibnamefont
  {M\"ockli}}\ and\ \bibinfo {author} {\bibfnamefont {M.}~\bibnamefont
  {Khodas}},\ }\bibfield  {title} {\bibinfo {title} {Robust parity-mixed
  superconductivity in disordered monolayer transition metal dichalcogenides},\
  }\href {https://doi.org/10.1103/PhysRevB.98.144518} {\bibfield  {journal}
  {\bibinfo  {journal} {Phys. Rev. B}\ }\textbf {\bibinfo {volume} {98}},\
  \bibinfo {pages} {144518} (\bibinfo {year} {2018})}\BibitemShut {NoStop}%
\bibitem [{\citenamefont {Robbins}\ \emph {et~al.}(2020)\citenamefont
  {Robbins}, \citenamefont {Annett},\ and\ \citenamefont
  {Gradhand}}]{PhysRevB.101.134505}%
  \BibitemOpen
  \bibfield  {author} {\bibinfo {author} {\bibfnamefont {J.}~\bibnamefont
  {Robbins}}, \bibinfo {author} {\bibfnamefont {J.~F.}\ \bibnamefont
  {Annett}},\ and\ \bibinfo {author} {\bibfnamefont {M.}~\bibnamefont
  {Gradhand}},\ }\bibfield  {title} {\bibinfo {title} {Theory of the orbital
  moment in a superconductor},\ }\href
  {https://doi.org/10.1103/PhysRevB.101.134505} {\bibfield  {journal} {\bibinfo
   {journal} {Phys. Rev. B}\ }\textbf {\bibinfo {volume} {101}},\ \bibinfo
  {pages} {134505} (\bibinfo {year} {2020})}\BibitemShut {NoStop}%
\end{thebibliography}%

\end{document}